\documentclass[preprint,preprintnumbers,amsmath,amssymb]{revtex4-1}
\usepackage{amsfonts}    % list packages between braces
\usepackage{amssymb}
\usepackage{amsmath}
\usepackage{latexsym}
\usepackage{eepic}
\usepackage{graphicx}
\usepackage[usenames,dvipsnames]{color}
\usepackage{color}

\usepackage[active]{srcltx}
\usepackage{epstopdf}

\newcommand{\al}{\alpha}
\newcommand{\pa}{\partial}
\newcommand{\ep}{\epsilon}
\newcommand{\vep}{\varepsilon}

\newcommand{\la}{\lambda}

\newcommand{\Om}{\Omega}
\newcommand{\om}{\omega}
\newcommand{\de}{\delta}

\newcommand{\De}{\Delta}

\newcommand{\rar}{\rightarrow}

\newcommand{\non}{\nonumber}

\begin{document}

\title{Three-body closed chain of interactive (an)harmonic oscillators and the algebra $sl(4)$}

\author{Alexander V Turbiner\\[6pt]
Instituto de Ciencias Nucleares, UNAM, M\'exico DF 04510, Mexico\\[4pt]
and\\
IHES, Bures-sur-Yvette, France\\[6pt]
turbiner@nucleares.unam.mx, turbiner@ihes.fr\\[6pt]
     Willard Miller, Jr.\\[6pt]
School of Mathematics, University of Minnesota, \\
Minneapolis, Minnesota, U.S.A.\\[6pt]
miller@ima.umn.edu\\
[8pt]
and \\[8pt]
M.A.~Escobar-Ruiz,\\[6pt]
%I nstituto de Ciencias Nucleares, UNAM, M\'exico DF 04510, Mexico\\
% and\\
Centre de Recherches Math\'ematiques, Universit\'e de Montr\'eal, \\
C.P. 6128, succ. Centre-Ville, Montr\'eal, QC H3C 3J7, Canada\\[6pt]
and\\
Departamento de F\'isica, UAM-I, M\'exico DF 09340, M\'exico\\[6pt]
escobarr@crm.umontreal.ca}

\begin{abstract}
In this work we study 2- and 3-body oscillators with quadratic and sextic pairwise potentials which depend on relative {\it distances}, $|{\bf r}_i - {\bf r}_j |$,  between particles.  The two-body harmonic oscillator is two-parametric and can be reduced to a one-dimensional radial Jacobi oscillator, while in the 3-body case such a reduction is not possible in general. Our study is restricted to solutions in the space of relative motion which are functions of mutual (relative) distances only ($S$-states). We pay special attention to the cases where the masses of the particles and spring constants are unequal as well as to the atomic, where one mass is infinite, and molecular, where two masses are infinite, limits.
In general, three-body harmonic oscillator is 7-parametric depending on 3 masses and 3 spring constants, and frequency. 
In particular, the first and second order integrals of the 3-body oscillator for unequal masses are searched: it is shown that for certain relations involving masses and spring constants the system becomes maximally (minimally) superintegrable in the case of two (one) relations.
\end{abstract}

\maketitle

\newpage

\section{Introduction}

The kinetic energy for the $n$-body quantum system of $d$-dimensional particles is of the form,
\begin{equation}
\label{Tflat}
   {\cal T}\ =\ -\sum_{i=1}^n \frac{1}{2m_i}\De_i^{(d)}\ ,
\end{equation}
with coordinate vector of $i$th particle ${\bf r}_i \equiv {\bf r}^{(d)}_i=(x_{i,1}\,,\cdots \,,x_{i,d})$ and mass $m_i$. Here, $\De_i^{(d)}$ is the $d$-dimensional Laplacian,
\[
     \De_i^{(d)}\ =\ \frac{\pa^2}{\pa{{\bf r}_i} \pa{{\bf r}_i}}\ ,
\]
associated with the $i$th particle. The quantum Hamiltonian of $n$-body problem has the form,
\begin{equation}
\label{Hamiltonian}
  {\cal H}\ =\ {\cal T}\ +\ V\ ,
\end{equation}
where the configuration space for ${\cal T}$ is ${\bf R}^{n \times d}$. The potential $V$ is translation-invariant.

The center-of-mass motion described by $d$-dimensional vectorial coordinate
\begin{equation}
\label{CMS}
    {\bf R}_{_0} \ =\ \frac{1}{\sqrt{M_n}}\,\sum_{k=1}^{n} m_k {\bf r}_{_k}\ ,\quad M_n={\sum_{j=1}^nm_j}\ ,
\end{equation}
can be separated out. After separation of the center-of-mass coordinate, the kinetic energy in the space of relative motion ${\bf R}_r \equiv {\bf R}^{(n-1)\times d }$ is described by
the flat-space Laplacian $\De_r^{(d(n-1))}$. Let $M_j = \sum_{k=1}^j m_k$, $j=1,\cdots,n-1$.
Remarkably, if the space of relative motion ${\bf R}_r$ is parameterized by $(n-1)$, $d$-dimensional, vectorial Jacobi coordinates
\begin{equation}
\label{Jacobi}
     {\bf r}^{(J)}_{j} \ = \ \sqrt{\frac{m_{j+1}M_j}{M_{j+1}}}\left({\bf r}_{j+1}-\sum_{k=1}^j\frac{m_k{\bf r}_k}{M_j}\right) ,
        \qquad\qquad j=1,\cdots,n-1 \  ,
\end{equation}
see e.g. \cite{Delves:1960} and also \cite{Willard:2018} for discussion, the flat-space, $(d (n-1))$-dimensional Laplacian of the relative motion becomes diagonal,
\begin{equation}
\label{Tflat-diag}
   {\cal T}\ =\ -\sum_{i=1}^n\frac{1}{m_i}\De_i^{(d)} \ = \ {\cal T}_0\  -  \
       \sum_{i=1}^{n-1}\frac{\pa^2}{\pa{{\bf r}_i^{(J)}} \pa{{\bf r}_i^{(J)}}}
       \equiv \ {\cal T}_0\ + {\cal T}_r\ ,
\end{equation}
where ${\cal T}_0=-\De_{{\bf R}_0}=-\frac{\pa^2}{\pa{{\bf R}_0} \pa{{\bf R}_0}}$ is the kinetic energy of the center-of-mass motion.
Note that the first Jacobi coordinate ${\bf r}^{(J)}_{1}$ is always proportional to the vector of relative distance between particles 1 and 2.
Evidently, the variables in ${\cal T}_r$ are separated and the kinetic energy of relative motion is the sum of kinetic energies in the Jacobi coordinate directions. By adding to ${\cal T}_r$ the harmonic
oscillator potentials in each Jacobi coordinate direction we arrive at the $n$-body harmonic oscillator as the collection of $(n-1)$ individual harmonic oscillators,
\begin{equation}
\label{H-osc-diag}
   {\cal H}_{r}\ =\   \
       \sum_{i=1}^{n-1} \bigg( \ -\frac{\pa^2}{\pa{{\bf r}_i^{(J)}} \pa{{\bf r}_i^{(J)}}}  + A_i\, \om^2\, { { ({\bf r}_i^{(J)}} } \cdot { {\bf r}_i^{(J)}})  \bigg)\ ,
\end{equation}
where $\om$ is the frequency and $A_i \geq 0,\ i=1, \ldots (n-1)$ are spring coefficients.
It is evident that this is an exactly-solvable problem: all eigenfunctions and eigenvalues are known analytically. It seems relevant to call this system the {\it Jacobi harmonic oscillator}. Let us note that if the potential in (\ref{Hamiltonian}), (\ref{H-osc-diag}) is chosen in the form of the moment of inertia $V=\sum^N_{i=1} m_i~{\bf r}_i^2$ all spring coefficients become equal to each other and also to a reduced mass of the system,
\[
      A_i\ =\ \mu \ \equiv\ \bigg(\frac{\prod_{j=1}^n m_j}{M}\bigg)^{\frac{1}{n-1}} \ ,\ M \equiv M_n \ ,
\]
see e.g. \cite{Delves:1960,Fortunato:2017}. Thus, in the space of vectorial Jacobi coordinates (\ref{Jacobi}) taking the moment of inertia as the potential leads to the isotropic Jacobi harmonic oscillator.
After the center-of-mass motion is removed,  the spectrum of (\ref{H-osc-diag}) is the sum of spectra of individual oscillators. Total zero angular momentum $L=0$ implies zero angular momenta of individual oscillators, hence the radial Hamiltonian of relative motion
is the sum of $d$-dimensional radial Hamiltonians,
\begin{equation}
\label{H-osc-diag-radial}
   {\cal H}_{r}^{(L=0)}\ =\
       \sum_{i=1}^{n-1} \bigg(\ -\frac{\pa^2}{\pa{{r}_i^{(J)}} \pa{{r}_i^{(J)}}} \ -\
       \frac{(d-1)}{{r}_i^{(J)}}\,\frac{\pa}{\pa{{r}_i^{(J)}}}
       \ +\ A_i\, \om^2\, { { ({r}_i^{(J)}} } \cdot { {r}_i^{(J)}})  \bigg)\ .
\end{equation}
Needless to say, the problem (\ref{H-osc-diag-radial}) is exactly-solvable (ES), its eigenfunctions are the product of individual eigenfunctions and the spectrum is linear in radial quantum numbers. Replacing the individual quadratic potential $A_j \om^2 (r^{(J)}_j)^2$ by the  quasi-exactly-solvable (QES) sextic potential, we arrive at the QES anharmonic Jacobi oscillator.
It is easy to check that in the $(n-1)$-dimensional space of relative radial motion of modules of Jacobi coordinates, or, saying differently, of the Jacobi distances ${r}_i^{(J)}$,  its hidden algebra is $sl_2^{\,\otimes \,{(n-1)}}$  acting on the  $(n-1)$-dimensional space, $\{{\rho}^{(J)}_{j}=|{\bf r}^{(J)}_{j}|^2\ , j=1, \ldots (n-1)\}$.

Since the spectrum of the Jacobi oscillators (\ref{H-osc-diag}), (\ref{H-osc-diag-radial}) is known explicitly, their eigenfunctions can be used as the basis to study many-body problems, as was proposed in \cite{Delves:1960}.
The present authors are not aware of any studies of the Jacobi oscillators {\it per se}.

In this work we will explore the case of 2- and 3-body oscillators with quadratic and sextic potentials which depend on relative {\it distances}, $|{\bf r}_i - {\bf r}_j |$, s.f. (\ref{Jacobi}), between particles.  The two-body harmonic oscillator, see Fig.1 for illustration, is reduced to a one-dimensional radial Jacobi oscillator, while in the 3-body case such a reduction is not possible in general.

\section{Two-body case}

At 1988 it was discovered that both the celebrated quantum one-dimensional harmonic oscillator and renown sextic Quasi-Exactly-Solvable (QES) anharmonic oscillator \cite{Turbiner-Ush:1987} possess the same hidden algebra $sl(2)$ \cite{Turbiner:1988} (for review see \cite{Turbiner:2016}). In different terms, this meant that for the two-body quantum problem with $d$ degrees of freedom of masses $m_1$ and $m_2$ there existed a (quasi)-exactly-solvable (sextic) quadratic potential in terms of relative distance $r_{12}$
for which finitely-many (infinitely-many) quantum $S$-states could be found by algebraic means.
Their eigenfunctions were the elements of the finite-dimensional representation space(s)
of $sl(2)$ algebra of differential operators.
This observation implied that, separating the center-of-mass (\ref{CMS}) and then making the change of variables  to the Euler coordinates in the space of relative motion,
\[
   ({\bf r}_1\ ,\ {\bf r}_2) \rar ({\bf R}_{0}\ ,\ \rho=r_{12}^2\ ,\ \Om)\ ,
\]
we wold arrive at the one-dimensional radial Schr\"odinger equation
\begin{equation}
\label{H-2-body}
       [-\De_{\rm rad}(\rho) \ + \ V(\rho) ]\,\Psi(\rho)\ =\ E\, \Psi(\rho)\ ,\qquad
    \De_{\rm rad}(\rho)\ =\ \frac{1}{\mu}
    \bigg( 2 \,\rho\, \pa^2_{\rho}\ +\ d\,\pa_{\rho} \bigg)\ ,
\end{equation}
where
\[
   H_{\rm rad}\ \equiv\ -\De_{\rm rad}\ + \ V\ ,
\]
is the radial Schr\"odinger operator, which governs one-dimensional (radial) dynamics, while
$\mu \ =\ \frac{m_1 \,m_2}{m_1+m_2}$ is the reduced mass. Note that in the atomic case,
when one of masses is large (or even infinitely large, $m_2 = \infty$)
the operator $\De_{\rm rad}$ remains in the {\it same} functional form, $\mu \rar m_1$.

\begin{figure}
  \centering
  \includegraphics[width=15.0cm]{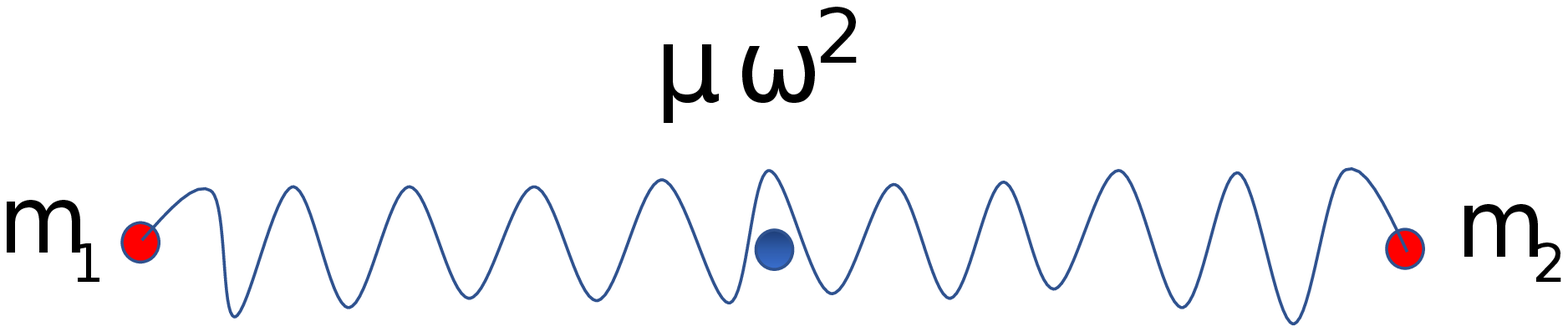}
  \caption{Two-body harmonic oscillator chain, Center-of-Mass marked by a blue bullet}
  \label{2B}
\end{figure}

The equation (\ref{H-2-body}) has finitely-many polynomial eigenfunctions for the quasi-exactly-solvable potential
\begin{equation}
\label{V2-qes}
   V^{(qes)}\ =\ 2\,\mu \,\bigg[\bigg(\om^2\, -\, A\, (4\,N +d+2)\bigg) \,\rho \ + \ 4\,\mu\, A\,\om\, \rho^2 \,+\, 4\,\mu^2\,A^2\, \rho^3\bigg]\ ,
\end{equation}
if $N$ is integer, and infinitely-many ones for the  exactly-solvable
\begin{equation}
\label{V2-exact}
   V^{(ex)}\ =\ 2\,\mu\,\om^2\, \rho\ ,
\end{equation}
potential. This  defines a 2-body QES anharmonic oscillator and a 2-body exactly-solvable (ES) harmonic oscillator, respectively, (for illustration see Fig.\ref{2B}). It is evident that both oscillators correspond to Jacobi (an)harmonic oscillators. In general, $A \geq 0$ and $\om > 0$ are parameters.

The ground state function for $N=0$ in (\ref{V2-qes}) is given by
\begin{equation}
\label{Psi0m-1d-qes}
   \Psi_0^{(qes)}\ =\ e^{-\, \mu\, \om\, \rho\, -\, \mu^2\, A\, \rho^2}\ ,
\end{equation}
with ground state energy $$E_0=\om d\ .$$ When $A=0$ the anharmonicity  disappears
and  $V^{(qes)}=V^{(es)}$, the expression (\ref{Psi0m-1d-qes}) becomes the ground
state function for the harmonic oscillator potential (\ref{V2-exact}).

Now on without loss of generality we put $m_1=m_2=1$, thus putting $\mu=1/2$.
Via a  gauge rotation, the radial Schr\"odinger operator $H_{rad}$ can be converted to the one-dimensional Hamiltonian,
\begin{equation}
\label{H-1d}
    {\cal H}_r \equiv \rho^{\frac{d-1}{4}}\, H_{\rm rad}\ \rho^{-\frac{d-1}{4}}\ =
    \ -\ \De_g(\rho) + V_{\rm eff} + V\ ,
\end{equation}
where the effective potential
\[
    V_{\rm eff}\ =\ \frac{(d-1)(d-3)}{4\, \rho}\ ,
\]
plays the role of a centrifugal force and $\De_g(\rho)=4\,\rho\, \pa^2_{\rho} + 2\, \pa_{\rho}$
is the Laplace-Beltrami operator with metric
\[
   g^{11}\ =\ 4\, \rho\ .
\]
If $d=1,3$ the effective potential vanishes, $V_{\rm eff}=0$. For $d=2$ the effective potential term is minimal.

It is evident that upon  changing $\rho$ to $r$: $r={\sqrt{\rho}}$ \,, the Laplace-Beltrami operator in variable $\rho$ becomes the second derivative in $r$,\ $\De_g=\pa^2_{r}$\,. Making de-quantization, i.e.,
replacing the quantum momentum (derivative) by the classical momentum,
\[
      -i\,\pa\ \rar\ P\ ,
\]
one can get a classical analogue of (\ref{H-1d}),
\begin{equation}
\label{HC-ld-rho}
   {H}^{(c)}_{\rm LB}(\rho)\ =\ g^{11}(\rho) P^2_{\rho} \ + \ V_{\rm eff} \ + \ V\ .
\end{equation}
It describes the internal motion of a one-dimensional body with coordinate dependent tensor of inertia $(g^{11})^{-1}$\, while the center of mass is kept fixed. In $r$-variables we arrive at
\begin{equation}
\label{HC-ld-r}
   {H}^{(c)}_{\rm LB}(r)\ =\ P^2_{r} \ + \ V_{\rm eff}\  + \ V\ ,
\end{equation}
which describes a one-dimensional classical (Q)ES (an)harmonic oscillator with centrifugal term if $d \neq 1,3$. Classical Hamiltonians (\ref{HC-ld-rho}) and (\ref{HC-ld-r}) are related through a contact canonical transformation,
\[
        \rho \,= \,r^2\ ,\quad  P_{\rho}\ = \ \frac{1}{2\, r}\, P_{r}\ .
\]
All trajectories for both potentials (\ref{V2-qes}), (\ref{V2-exact}) are periodic,
and both QES and ES periods can be easily found.

The QES radial Schr\"odinger operator $H_{\rm rad}$ with the potential (\ref{V2-qes}) can be converted into the algebraic operator by making the gauge rotation,
\begin{equation}
\label{h-1d-qes}
\begin{aligned}
   h^{(qes)}(\rho) \ & \equiv \ (\Psi_0^{(qes)})^{-1}\, (H_{\rm rad}-E_0)\, \Psi_0^{(qes)}
   \\ &
    =\ -4\,\rho\,\pa_\rho^2\,+\,2\,\left(2\,A\,\rho^2\,+\,2\,\om\,\rho\,-\,d \right)\pa_\rho
    \,-\,4\,A\,N\, \rho \ ,
\end{aligned}
\end{equation}
where
\begin{equation}
\label{Psi0-1d-qes}
   \Psi_0^{(qes)}\ =\  e^{-\frac{\om}{2}\rho\,-\,\frac{A }{4}\rho^2}\ ,
\end{equation}
c.f. (\ref{Psi0m-1d-qes}),
\begin{equation}
\label{E0}
   E_0\ =\ d\,\om\ .
\end{equation}
One can check that the operator $h^{(qes)}$ has a single finite-dimensional invariant subspace
\begin{equation}
\label{P1-rho}
     {\mathcal P}_{N}\ \equiv \ \langle \rho^{p}\ \vert \
     0 \le p \le N \rangle\ ,
\end{equation}
which coincides with the finite-dimensional representation space of the Lie algebra $sl(2)$ of the first order differential operators
\begin{equation}
\label{sl2}
    {\cal J}^+(N)\ =\ \rho^2\,\pa_{\rho} - N \rho\ ,\qquad {\cal J}^0(N)\ =\ \rho\,\pa_{\rho} - N\ ,\qquad {\cal J}^-\ =\ \pa_{\rho}\ .
\end{equation}
Thus, the operator $h^{(qes)}$ (\ref{h-1d-qes}) can be written in terms of $sl(2)$ algebra generators,
\begin{equation}
\label{h-1d-qes-Lie}
    h^{(qes)}\ =\   -4\,{\cal J}^0\,{\cal J}^- +4\,A\,{\cal J}^+ \ - \ 2\,(d+2\,N)\,{\cal J}^- \ + \ 4\,\om\,{\cal J}^0 \ + \ 4\,N\,\om     \ ,
\end{equation}
Putting $A=0$ in (\ref{h-1d-qes-Lie}) we arrive at the exactly-solvable operator
\begin{equation}
\label{h-1d-es-rho}
    h^{(es)}(\rho)\ \equiv \ (\Psi_0^{(es)})^{-1}\,(H_{\rm rad}-E_0)\, \Psi_0^{(es)}\ =\
    -4\,\rho\, \pa_\rho^2
\  +\ 2 \,\left(2\, \rho\, \om - d \right)\pa_\rho\ ,
\end{equation}
where
\begin{equation}
\label{Psi0-1d-es}
   \Psi_0^{(es)}\ =\  e^{-\frac{1}{2}\om \rho}\ ,
\end{equation}
is the ground state function and $E_0$ is given by (\ref{E0}).
Thus, the operator $h^{(es)}$ can be written in terms of $sl(2)$ generators
${\cal J}^0, {\cal J}^-$, see (\ref{sl2}),
\begin{equation}
\label{h-1d-es-Lie}
    h^{(es)}\ =\     -4\,{\cal J}^0\,{\cal J}^-  \ - \ 2\,(d+2\,N)\,{\cal J}^- \ + \ 4\,\om\,{\cal J}^0 \ + \ 4\,N\,\om\ .
\end{equation}
It can be immediately recognized that the operator $h^{(es)}$ (\ref{h-1d-es-rho}) is the Laguerre operator. Hence, the spectral problem,
\[
  h^{(es)} \,\phi \ =\ \ep \,\phi \ ,
\]
has infinitely-many polynomial eigenfunctions,
\[
    \phi_n \ = \ L_n^{(\frac{d}{2}-1)}(\om \rho)\ ,\ n\,=\,0,1,2, \ldots \ ,
\]
which are the Laguerre polynomials $L_n^{(\al)}$ with equidistant spectra
\[
    \ep_n\ =\ 4\, \om\, n\ .
\]

\section{Three-Body Case}

The general quantum Hamiltonian for three $d$-dimensional ($d>1$) bodies of masses $m_1, m_2, m_3$
with translation-invariant potential, which depends on relative (mutual) distances
between particles only, is of the form,
\begin{equation}
\label{Hgen}
   {\cal H}\ =\ -\sum_{i=1}^3 \frac{1}{2 \,m_i} \De_i^{(d)}\ +\  V(r_{12},\,r_{13},\,r_{23})\ ,\
\end{equation}
see e.g. \cite{TMA:2016,TME3-d}, where $\De_i^{(d)}$ is $d$-dimensional Laplacian of $i$th particle with coordinate vector ${\bf r}_i \equiv {\bf r}^{(d)}_i=(x_{i,1}\,, x_{i,2}\,,x_{i,3}\ldots \,,x_{i,d})$\ , and
\begin{equation}
\label{rel-coord}
r_{ij}=|{\bf r}_i - {\bf r}_j|\ ,\quad i,j=1,2,3\ ,
\end{equation}
is the (relative) distance between particles $i$ and $j$, $r_{ij}=r_{ji}$. Separating the center-of-mass (\ref{CMS}) and then making the change of variables in the space of relative motion to  generalized Euler coordinates (three relative distances (\ref{rel-coord}) and $(2d-3)$ angles)
\[
   ({\bf r}_1\ ,\ {\bf r}_2\ ,\ {\bf r}_3)\ \rar \ ({\bf R}_{0}\ ,\ \rho_{12}=r_{12}^2\ ,\ \rho_{13}=r_{13}^2\ , \ \rho_{23}=r_{23}^2\ , \  \Om)\ ,
\]
we arrive at a three-dimensional radial-type Schr\"odinger equation \cite{TME3-d}
\begin{equation}
\label{H-3-body}
       [-\De_{\rm rad}(\rho)\ + \ V(\rho)]\,\Psi(\rho)\ =\ E \,\Psi(\rho)\ ,\
\end{equation}
where
\[
  \frac{1}{2}\De_{\rm rad}(\rho)\ =\
  \frac{1}{\mu_{12}} \rho_{12}\, \pa_{\rho_{12}}^2 +
  \frac{1}{\mu_{13}} \rho_{13}\, \pa_{\rho_{13}}^2 +
  \frac{1}{\mu_{23}} \rho_{23}\, \pa_{\rho_{23}}^2 +
\]
\[
  \frac{(\rho_{13} + \rho_{12} - \rho_{23})}{m_1}\pa_{\rho_{13},\,\rho_{12}}\ + \
  \frac{(\rho_{13} + \rho_{23} - \rho_{12})}{m_3}\pa_{\rho_{13},\,\rho_{23}}\ + \
  \frac{(\rho_{23} + \rho_{12} - \rho_{13})}{m_2}\pa_{\rho_{23},\,\rho_{12}}\ + \
\]
\begin{equation}
\label{addition3-3r-M}
 \frac{d}{2}\,\bigg[ \frac{1}{\mu_{12}} \pa_{\rho_{12}} +
  \frac{1}{\mu_{13}} \pa_{\rho_{13}} +
  \frac{1}{\mu_{23}} \pa_{\rho_{23}} \bigg] \ ,
\end{equation}
c.f. \cite{TMA:2016}, and
\[
   \frac{1}{\mu_{ij}}\ = \ \frac{m_i+m_j}{m_i\, m_j}\ ,
\]
is the inverse reduced mass, which governs three-dimensional (radial) dynamics in variables $r_{12}, r_{13}, r_{23}$. The operator
\begin{equation}
\label{Hrad3}
  H_{\rm rad}\ \equiv\ -\De_{\rm rad} \ + \  V\ ,
\end{equation}
is, in fact, the three-dimensional, radial Schr\"odinger operator, see \cite{TME3-d}. It can be called three-dimensional radial Hamiltonian. One can show that for the potential
\begin{equation}
\label{V3-es}
   V^{(ex)}\ =\ 2\,\om^2\bigg[   \nu_{12}\,\rho_{12} \ + \  \nu_{13}\,\rho_{13} \ + \  \nu_{23}\,\rho_{23} \bigg]\ ,
\end{equation}
the equation (\ref{H-3-body}) has infinitely-many polynomial eigenfunctions,
where $\om > 0$ and $\nu_{12},\,\nu_{13},\,\nu_{23}$ are certain positive mass-dependent parameters, see below. Note that as for the one-dimensional case $d=1$, the 3-body problem with potential (\ref{V3-es}) was analyzed in \cite{Fernandez}. Here, we consider the general case $d>1$.

\begin{figure}
  \centering
  \includegraphics[width=15.0cm]{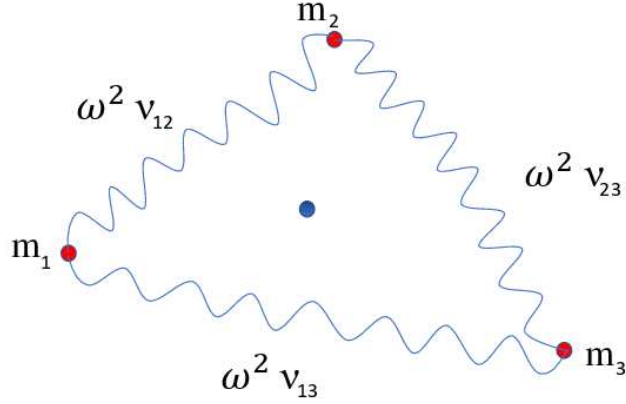}
  \caption{Three-body harmonic oscillator chain}
  \label{3-body}
\end{figure}

A further remark is in order. The potential (\ref{V3-es}) is nothing but a three-dimensional anisotropic oscillator in $\rho$-variables, hence, in the space of relative motion, see Fig.\ref{3-body}. The corresponding configuration space (the physics domain) is confined to the cube ${\bf R}_+(\rho_{12}) \times {\bf R}_+(\rho_{13}) \times {\bf R}_+(\rho_{23})$ in $E_3$. More explicitly, it is given by the condition
\[
  2(\,\rho_{12}\,\rho_{13}\ + \ \rho_{12}\,\rho_{23}\ + \ \rho_{23}\,\rho_{13})\ - \ (\rho_{12}^2\ + \
  \rho_{13}^2 \  + \ \rho_{23}^2)\ \geq \ 0 \ ,
\]
\noindent
stating that the square of the area of the triangle formed by the particle positions must be greater or equal than zero.

Now, it is easy to check that the ground state function in (\ref{Hrad3}), (\ref{V3-es}) is given by
\begin{equation}
\label{Psi03}
   \Psi_0^{(es)}\ =\ e^{-\om\, (a\,\mu_{12}\,\rho_{12}\,+\,b\,\mu_{13}\,\rho_{13}\,+\,c\,\mu_{23}\,\rho_{23})}\ ,
\end{equation}
where $a,\,b,\,c \geq 0$ have the meaning of spring constants, with the ground state energy
\begin{equation}
\label{E0en}
E_0\ = \ \om \,d\,(a+b+c) \ ,
\end{equation}
which is mass-independent. Here
\begin{equation}
\label{freq-3}
\begin{aligned}
& \nu_{12} \ = \  a^2\, \mu _{12} \ + \ a\, b\ \frac{\mu _{12}\, \mu _{13} }{m_1} \ + \  a\, c\ \frac{\mu _{12} \,\mu _{23} }{m_2} \ - \  b\, c \ \frac{\mu_{13} \,\mu _{23} }{m_3} \ ,
\\ &
\nu_{13} \ = \  b^2\, \mu _{13} \ + \  a\, b\ \frac{\mu _{12} \,\mu _{13} }{m_1}\ + \  b \,c\ \frac{\mu _{13} \,\mu _{23}}{m_3}\  - \ a\, c\ \frac{\mu _{12} \,\mu _{23} }{m_2} \ ,
\\ &
\nu_{23} \ = \ c^2 \,\mu _{23} \ + \  a \,c \ \frac{\mu _{12} \,\mu _{23} }{m_2}\ + \  b\, c\ \frac{\mu _{13} \,\mu _{23} }{m_3} \ - \   a\, b\ \frac{\mu _{12}\, \mu _{13} }{m_1} \ .
\end{aligned}
\end{equation}

After a gauge rotation of (\ref{Hrad3}) with gauge factor
\begin{equation}
\label{Gamma}
  \Gamma \ = \ \bigg[\frac{(2\,\rho_{12}\,\rho_{13}+2\,\rho_{12}\,\rho_{23}+2\,\rho_{23}\,\rho_{13}-\rho_{12}^2-
  \rho_{13}^2-\rho_{23}^2)^{2-d}}{ {m_1 m_2\, \rho_{12}+m_1 m_3\, \rho_{13}+m_2 m_3 \,\rho_{23}} }\bigg]^{\frac{1}{4}} \ ,
\end{equation}
the radial Schr\"odinger operator $H_{\rm rad}$ (\ref{Hrad3}) is converted to a three-dimensional one-particle Hamiltonian \cite{TME3-d},
\begin{equation}
\label{H-3body-one-particle}
   {\cal H}_r \equiv   \Gamma^{-1}\,  [-\De_{\rm rad}(\rho) + V(\rho)]\,\Gamma  \ =\  -\De_{\rm LB}(\rho)\ + \ V(\rho)\  + \ V^{(\rm eff)}\ ,
\end{equation}
where $\De_{\rm LB}$ is the Laplace-Beltrami operator
\[
   \De_{LB}(\rho)\ =\ {\sqrt{\rm D}}\ \pa_{\mu}\ \frac{1}{\sqrt{\rm D}}\  g^{\mu \nu} \pa_{\nu}\ , \qquad \nu,\mu=1,2,3 \ ,
\]
and  $\pa_{1}\equiv \pa_{\rho_{12}},\pa_{2}\equiv \pa_{\rho_{13}} \pa_{3}\equiv\pa_{\rho_{23}}$,
with the co-metric
\begin{equation}
\label{gmn33-rho}
 g^{\mu \nu}(\rho)\ = \left|
 \begin{array}{ccc}
 \frac{2}{\mu_{12}} \rho_{12} & \ \frac{(\rho_{13} + \rho_{12} - \rho_{23})}{m_1} & \ \frac{(\rho_{23} + \rho_{12} - \rho_{13})}{m_2} \\
            &                                   &                                   \\
 \frac{(\rho_{13} + \rho_{12} - \rho_{23})}{m_1} & \  \frac{2}{\mu_{13}} \rho_{13} & \ \frac{(\rho_{13} + \rho_{23} - \rho_{12})}{m_3} \\
            &  \                                  &                                   \\
 \frac{(\rho_{23} + \rho_{12} - \rho_{13})}{m_2} & \ \frac{(\rho_{13} + \rho_{23} - \rho_{12})}{m_3} & \frac{2}{\mu_{23}} \rho_{23}
 \end{array}
               \right| \ .
\end{equation}
Its determinant
\[
  D\ \equiv \ {\rm Det} g^{\mu \nu}\ =\ 2\,\frac{m_1+m_2+m_3}{m_1^2m_2^2m_3^2} \times
\]
\begin{equation}
\label{gmn33-rho-det-M}
 \left(m_1m_2\,\rho_{12}+m_1m_3\,\rho_{13}+m_2m_3\,\rho_{23}\right)
                     \left(2\,\rho_{12}\,\rho_{13} + 2\, \rho_{12}\,\rho_{23} + 2\, \rho_{13}\,\rho_{23}-\rho_{12}^2- \rho_{13}^2 - \rho_{23}^2\right) \ ,
\end{equation}
is in factorisable form and, in general, is positive definite, while $V^{(\rm eff)}$ is the effective potential
\begin{equation}
\label{Veff}
V^{(\rm eff)} \ =\ \frac{3}{8}\ \frac{(m_1+m_2+m_3)}{\left(m_1 m_2 \,\rho_{12}+m_1 m_3 \,\rho_{13}+m_2 m_3 \,\rho_{23}\right)}\ +
\end{equation}
\[
 +\ \frac{(d-2)(d-4)}{2}\ \frac{\left(m_1 m_2 \rho_{12}+m_1 m_3 \rho_{13}+m_2 m_3 \rho_{23}\right)}
  { m_1 m_2 m_3
  \left( 2\,\rho_{12}\,\rho_{13}+2\,\rho_{12}\,\rho_{23}+2\,\rho_{23}\,\rho_{13}-\rho_{12}^2-
  \rho_{13}^2-\rho_{23}^2
  \right)}\ ,
\]
which plays the role of a centrifugal potential. The second term vanishes at $d=2,4$. Furthermore at $d=3$ the effective potential is the smallest function: the second term becomes negative. It must be emphasized that the mass-independent expression
$( 2\,\rho_{12}\,\rho_{13}+2\,\rho_{12}\,\rho_{23}+2\,\rho_{23}\,\rho_{13}-\rho_{12}^2-   \rho_{13}^2-\rho_{23}^2)$, which enters to (\ref{Gamma}), (\ref{gmn33-rho-det-M}), (\ref{Veff}), is the square of the area of the triangle formed by
particle positions, see Fig.1,
\[
    16\,S_{\triangle}^2\ =\  2\,\rho_{12}\,\rho_{13}+2\,\rho_{12}\,\rho_{23}+2\,\rho_{23}\,\rho_{13}-\rho_{12}^2-
  \rho_{13}^2-\rho_{23}^2 \ ,
\]
\cite{TME3-d}.
By making another gauge rotation with $\Psi_0^{(es)}$ (\ref{Psi03}) as the gauge factor, we convert  the radial Schr\"odinger operator $H_{\rm rad}$ (\ref{Hrad3}) to an algebraic operator, where the coefficient functions are polynomials,
\[
 h^{(ex)} \equiv \big( \Psi_0^{(es)}\big)^{-1}\, [-\De_{\rm rad}(\rho) + V(\rho) - E_0]\, \Psi_0^{(es)}
 \ =
\]
\[
     -2\,\bigg[  \frac{1}{\mu_{12}} \rho_{12}\, \pa_{\rho_{12}}^2\ + \
     \frac{1}{\mu_{13}} \rho_{13}\, \pa_{\rho_{13}}^2\ +\
       \frac{1}{\mu_{23}} \rho_{23}\, \pa_{\rho_{23}}^2\ +
\]
\begin{equation}
\label{H-3body-algebraic}
\begin{aligned}
&  \frac{(\rho_{13} + \rho_{12} - \rho_{23})}{m_1}\pa_{\rho_{13},\,\rho_{12}} +
   \frac{(\rho_{23} + \rho_{12} - \rho_{13})}{m_2}\pa_{\rho_{23},\,\rho_{12}} +
  \frac{(\rho_{13} + \rho_{23} - \rho_{12})}{m_3}\pa_{\rho_{13},\,\rho_{23}}  \bigg]\ +
\\ &
% \bigg[
 \frac{ 2\mu _{12} \omega  \left( 2 a m_1 m_2 \rho _{12}+b \mu _{13} m_2 \left(\rho _{12}+\rho _{13}-\rho _{23}\right)+c \mu _{23} m_1 \left(\rho _{12}+\rho _{23}-\rho _{13}\right)\right)-d m_1 m_2}{\mu _{12}\, m_1 m_2}   \,\pa_{\rho_{12}}\ +
\\ &
 \frac{ 2\mu _{13} \omega  \left( 2 b m_1 m_3 \rho _{13}  +  a \mu _{12} m_3 \left(\rho _{12}+\rho _{13}-\rho _{23}\right)+c \mu _{23} m_1 \left(\rho _{13}+\rho _{23}-\rho _{12}\right)\right)-d m_1 m_3}{\mu _{13}\, m_1 m_3}\,\pa_{\rho_{13}}\ +
\\ &
\frac{ 2\mu _{23} \omega  \left(2 c m_2 m_3 \rho _{23} +  a \mu _{12} m_3 \left(\rho _{12}+\rho _{23}-\rho _{13}\right)+b \mu _{13} m_2 \left(\rho _{13}+\rho _{23}-\rho _{12}\right)\right)-d m_2 m_3}{\mu _{23}\, m_2 m_3}\,\pa_{\rho_{23}}\ .
%\bigg] \ ,
\end{aligned}
\end{equation}
Here $E_0$ is given by (\ref{E0en}). It is easy to check that (\ref{H-3body-algebraic}) preserves the triangular space of
polynomials
\begin{equation}
\label{P3}
     {\mathcal P}^{(1,1,1)}_{N}\ =\ \langle \rho_{12}^{p_1}\, \rho_{13}^{p_2} \,\rho_{23}^{p_3} \vert \
     0 \le p_1 + p_2+ p_3 \le N \rangle\ ,\ N=0,1,2, \ldots\ .
\end{equation}
for any integer $N$. Hence, it preserves the flag ${\mathcal P}^{(1,1,1)}$ with the characteristic (weight) vector $(1,1,1)$.

Note that the operator (\ref{H-3body-algebraic}) is of Lie-algebraic nature: it can be rewritten in terms of the generators  of the maximal affine sub-algebra  $b_4$ of the algebra $sl(4,{\bf R})$ realized by the first order differential operators, see e.g. \cite{Turbiner:1988,Turbiner:2016}
\begin{eqnarray}
\label{sl4R}
 {\cal J}_i^- \ &=& \ \frac{\pa}{\pa u_i}\ ,\qquad \quad i=1,2,3\ , \non  \\
 {{\cal J}^0_{ij}} \ &=& \
               u_i \frac{\pa}{\pa u_j}\ , \qquad i,j=1,2,3 \ ,\non \\
 {\cal J}^0(N)\  &=& \ \sum_{i=1}^{3} u_i \frac{\pa}{\pa u_i}-N\, , \non \\
 {\cal J}_i^+(N) \ &=& \ u_i \,{\cal J}^0(N)\ =\
    u_i\, \left( \sum_{j=1}^{3} u_j\frac{\pa}{\pa u_j}-N \right)\ ,
       \quad i=1,2,3\ ,
\end{eqnarray}
where $N$ is a parameter and it is denoted
\[
 u_1\equiv\rho_{12}\ ,\qquad u_2\equiv\rho_{13}\ , \qquad u_3\equiv\rho_{23} \ .
\]
If $N$ is a non-negative integer, a finite-dimensional representation space appears,
\begin{equation}
\label{P3-sl4}
     {\cal P}_{N}\ =\ \langle u_1^{p_1}\, u_2^{p_2}\, u_3^{p_3} \vert \ 0 \le p_1+p_2+p_3 \le N \rangle\ .
\end{equation}
This space coincides with (\ref{P3}). It is easy to check that the space ${\cal P}_{N}$ is invariant with respect to $3D$ projective transformations,
\[
   u_i \rar \frac{a_i u_1 + b_i u_2 + c_i u_3 + d_i}
                 {\al u_1 + \beta u_2 + \gamma u_3 + \delta}\ ,\quad i=1,2,3 \ ,
\]
where $a_i, b_i, c_i, d_i, \al , \beta, \gamma, \de$ are real parameters. By taking the parameters $a, b, c$'s at $i=1,2,3$  and $\al, \beta, \gamma, \de$ as the rows of the 4 x 4 matrix $G$ one can demonstrate that $G \in GL(4,R)$.

The spectrum of (\ref{H-3body-algebraic}) depends on four integers (quantum numbers) and is linear in quantum numbers. In terms of the generators (\ref{sl4R}), the algebraic operator (\ref{H-3body-algebraic}) takes the form

\[
 h^{(ex)}({\cal J}) \ = \
     -2\,\bigg[  \frac{1}{\mu_{12}} {{\cal J}^0_{11}}\,{{\cal J}^-_{1}} \ + \
     \frac{1}{\mu_{13}}  {{\cal J}^0_{22}}\,{{\cal J}^-_{2}}  \ +\
       \frac{1}{\mu_{23}}  {{\cal J}^0_{33}}\,{{\cal J}^-_{3}}  \ + \ \frac{1}{m_1}({{\cal J}^0_{22}}\,{{\cal J}^-_{1}}  +  {{\cal J}^0_{11}}\,{{\cal J}^-_{2}}  - {{\cal J}^0_{31}}\,{{\cal J}^-_{2}})\ +
\]
\begin{equation}
\label{H-3body-Lie-algebraic}
\begin{aligned}
&
  \frac{1}{m_2}({{\cal J}^0_{33}}\,{{\cal J}^-_{1}}  +  {{\cal J}^0_{11}}\,{{\cal J}^-_{3}}  - {{\cal J}^0_{23}}\,{{\cal J}^-_{1}}) \ + \
  \frac{1}{m_3}({{\cal J}^0_{22}}\,{{\cal J}^-_{3}}  +  {{\cal J}^0_{33}}\,{{\cal J}^-_{2}}  - {{\cal J}^0_{12}}\,{{\cal J}^-_{3}})\,\bigg]\ +
\\ &
% \bigg[
 \frac{2\,\mu_{12}\,\om \bigg(
 (2 a m_1 m_2+ b \mu _{13} m_2+c \mu_{23} m_1) {{\cal J}^0_{11}} +
(b \mu_{13} m_2 - c \mu_{23} m_1) ({{\cal J}^0_{21}}-{{\cal J}^0_{31}})
 \bigg)}{\mu_{12}\, m_1\, m_2}  \ +
\\ &
 \frac{ 2\,\mu _{13}\, \om \bigg(
 (2 b m_1 m_3 + a \mu _{12} m_3 + c \mu_{23} m_1) {{\cal J}^0_{22}} +
(a \mu_{12} m_3 - c \mu_{23} m_1) ({{\cal J}^0_{12}}-{{\cal J}^0_{32}})
 \bigg)}{\mu _{13}\, m_1\, m_3}\ +
\\ &
\frac{ 2\,\mu _{23}\, \om  \bigg(
(2 c m_2 m_3 + a \mu _{12} m_3 + c \mu_{13} m_2) {{\cal J}^0_{33}} +
(a \mu_{12} m_3 - b \mu_{13} m_2) ({{\cal J}^0_{13}}-{{\cal J}^0_{23}})
 \bigg)}{\mu _{23}\, m_2\, m_3} \ -
\end{aligned}
\end{equation}
\[
 -\ d \,\left(\frac{{\cal J}^-_{1}}{\mu _{12}}\ -\ \frac{{\cal J}^-_{2}}{\mu _{13}}\ -\   \frac{{\cal J}^-_{3}}{\mu _{23}}\right)\ .
\]

It is worth mentioning that for arbitrary masses $m_1,m_2,m_3$ the Hamiltonian
\begin{equation}
\label{Hp}
 \tilde{\cal H}\ \equiv \ -\De_{\rm rad}(\rho)\  + \ V^{(ex)}(\rho) \ + \ \tilde{V}(\rho) \ ,
\end{equation}
with the cubic in $\rho$'s potential
\[
\tilde{V}(\rho) \ = \  8\,\bigg[\frac{ A_{12}^2 }{\mu _{12} }\,\rho _{12}^3 \,+\,\frac{ A_{13}^2}{\mu _{13} }\,\rho _{13}^3 \ +\ \frac{ A_{23}^2 }{\mu _{23} }\,\rho _{23}^3  \ +
\]
\[
  A_{12} \left(\frac{ A_{13}}{m_1}\rho _{13}^2\, + \, \frac{ A_{23}}{m_2}\rho _{23}^2\right)\,
\rho _{12} \, + \, A_{13} \left(\frac{ A_{12}}{m_1}\rho _{12}^2\, + \, \frac{ A_{23}}{m_3}\rho _{23}^2\right)\,\rho _{13}\ + \ A_{23} \left(\frac{ A_{12}  }{m_2}\rho _{12}^2+\frac{ A_{13} }{m_3}
\rho _{13}^2\right)\,\rho _{23} \ -
\]
\[
 \bigg(\frac{ A_{12} \,A_{13} }{m_1}\ + \ \frac{ A_{12} \,A_{23}}{m_2} \  + \
 \frac{ A_{13} \,A_{23} }{m_3}\bigg)\,\rho _{12}\, \rho _{13}\, \rho _{23}  \bigg]
\ + \
\]
\[
     \frac{4\,\om}{m_1\,m_2\,m_3}\,\bigg
     [{A_{12}\,m_3\left(2\,a\,m_1\,m_2\,+\,b\,\mu_{13}\,m_2\,+\,c\,\mu_{23}\,m_1\right)}
    \,\rho_{12}^2\ +
\]
\[
    {A_{13}\,m_2\left(a\,\mu_{12}\,m_3\,+\,2\, b\, m_1 \,m_3\,+\,c \,\mu_{23}\,m_1\right)}
    \,\rho_{13}^2\ +
\]
\[
    {A_{23}\,m_1\left(a\,\mu_{12}\,m_3\,+\,b\,\mu_{13}\,m_2\,+\,2\,c\,m_2\,m_3\right)}
    \,\rho_{23}^2\ +
\]
\[
 {\left(A_{13}\,m_2\,\left(a \mu_{12} m_3-c \mu _{23} m_1\right)\ +\ A_{12} m_3
  \left(b \mu_{13} m_2-c \mu_{23} m_1\right)\right)}\,\rho_{12}\,\rho_{13}\ +
\]
\[
 {\left(A_{23}\,m_1\,\left(a \,\mu_{12}\, m_3-b \,\mu_{13} \,m_2\right)\ +\ A_{12}\, m_3\, \left(c\, \mu_{23}\, m_1-b \,\mu_{13}\,m_2\right)\right)} \,\rho_{12}\,\rho_{23}\ +
\]
\[
  {\left(A_{23}\,m_1\,\left(\,b \mu_{13} m_2 - a \mu_{12} m_3\,\right)\ +\
   A_{13}\,m_2\,\left(\,c \mu_{23} m_1-a \mu_{12} m_3\,\right)\right)}
   \,\rho_{13}\,\rho_{23}\,\bigg]\ -
\]
\begin{equation}
\label{QES-privitive}
   2\,(d+2)\,\bigg[\frac{ A_{12} }{\mu _{12}}\rho_{12} \ + \ \frac{A_{13}}{\mu_{13}}\rho_{13}
        \ + \ \frac{A_{23}}{\mu_{23}}\rho_{23}\,\bigg]\ ,
\end{equation}
corresponds to a primitive QES problem. Here only the ground state function
\begin{equation}
\label{psiQES-3}
 \tilde {\Psi}_0 \ =\ \Psi_0^{(ex)} \times e^{-(A_{12}\,\rho_{12}^2+A_{13}\,\rho_{13}^2+A_{23}\,\rho_{23}^2)} \ ,
\end{equation}
is known explicitly, with  constants $A_{12}, A_{13}, A_{23} \geq 0$.
The operator (\ref{Hp}) has no invariant subspaces except for $<1>$. We are unable to find other QES problems.

For the 3-body harmonic oscillator with potential (\ref{V3-es}) there are three important physically particular cases defined by values of masses:
(i) the case of three equal masses,
(ii) atomic like case, when one mass is infinite,
(iii) molecular like case when two masses are infinite.
and also (iv) the one-dimensional case $d=1$. They will be studied in detail.

\subsection{Three particles of equal masses}

\subsubsection{ {\it Arbitrary $a,b,c$}}

Let us take the eigenvalue problem (\ref{H-3-body}) with potential (\ref{V3-es}) and consider the case of three particles of equal masses, namely $m_1=m_2=m_3=m$, but different spring constants $a,b,c>0$. The exactly solvable potential (\ref{V3-es}) becomes
\begin{equation}
\label{V3-es-meq}
   V^{(3m)}\ =\ \frac{1}{2}\,m\,\om^2\,[\,(2 a^2+a (b+c)-b c)\,\rho_{12} \ + \ (2b^2+b(a+c)-a c)\,\rho_{13} \ + \ (2c^2+c(a+b)-a b)\,\rho_{23}\,] \ ,
\end{equation}
c.f. (\ref{freq-3}). This is a type of non-isotropic 3-body harmonic oscillator with different spring constants.
In this case the ground state function (\ref{Psi03}) is reduced to
\begin{equation}
\label{Psi03-meq}
   \Psi_0^{(3m)}\ =\ e^{-\frac{\om\,m }{2}\,(\,a\,\rho_{12}\ +\ b\,\rho_{13}\ +\ c\,\rho_{23}\,)}\ ,
\end{equation}
while its energy (\ref{E0en}) remains unchanged
\begin{equation*}
\label{e03-meq}
E_0^{(3m)}\ = \ \om \,d\,(a+b+c) \ .
\end{equation*}
The algebraic radial Schr\"odinger operator (\ref{H-3body-algebraic}) simplifies to
\[
 h^{(ex)}(\rho) \ = \
     -\frac{2}{m}\,\bigg[  2\, (\,\rho_{12}\, \pa_{\rho_{12}}^2\ + \
    \rho_{13}\, \pa_{\rho_{13}}^2\ +\
       \rho_{23}\, \pa_{\rho_{23}}^2\,)
\]
\begin{equation}
\label{H-3body-algebraic-meq}
\begin{aligned}
& \ + \ {(\rho_{13} + \rho_{12} - \rho_{23})}\pa_{\rho_{13},\,\rho_{12}}\ + \
   {(\rho_{23} + \rho_{12} - \rho_{13})}\pa_{\rho_{23},\,\rho_{12}}\ + \
  {(\rho_{13} + \rho_{23} - \rho_{12})}\pa_{\rho_{13},\,\rho_{23}}  \bigg]
\\ &
% \bigg[
  \hskip 1cm + \ \om \,\bigg[\,(4a+b+c)\,\rho_{12}\,\pa_{\rho_{12}} + (4b+a+c)\,\rho_{13}\pa_{\rho_{13}}
   + (4c+a+b)\,\rho_{23}\pa_{\rho_{23}}\ +\
\\ &
   \hskip 1cm (b-c) \, \left(\rho _{13}-
   \rho_{23}\right)\pa_{\rho_{12}}\ +\
   (a - c)\, \left(\rho_{12}-\rho _{23}\right)\pa_{\rho_{13}}\ + \ (a-b) \, \left(\rho _{12}-\rho _{13}\right)\pa_{\rho_{23}}\,\bigg]
\\ &
   \hskip 5cm - \  \frac{2\,d}{ m}\,\bigg[\pa_{\rho_{12}}\ +\ \pa_{\rho_{13}}\ +\ \pa_{\rho_{23}}\bigg] \ ,
\end{aligned}
\end{equation}
as well as the corresponding Lie-algebraic operator (\ref{H-3body-Lie-algebraic})
\[
 h^{(ex)}({\cal J}) \ = \
     -\frac{2}{m}\,\bigg[  2\, ({{\cal J}^0_{11}}\,{{\cal J}^-_{1}} \ + \
       {{\cal J}^0_{22}}\,{{\cal J}^-_{2}}  \ +\
        {{\cal J}^0_{33}}\,{{\cal J}^-_{3}} ) \ + \ {{\cal J}^0_{22}}\,{{\cal J}^-_{1}}  +  {{\cal J}^0_{11}}\,{{\cal J}^-_{2}}  - {{\cal J}^0_{31}}\,{{\cal J}^-_{2}}\ +
\]
\begin{equation}
\label{H-3body-Lie-algebraic-meq}
\begin{aligned}
&
  {{\cal J}^0_{33}}\,{{\cal J}^-_{1}}  +  {{\cal J}^0_{11}}\,{{\cal J}^-_{3}}  - {{\cal J}^0_{23}}\,{{\cal J}^-_{1}} \ + \
 {{\cal J}^0_{22}}\,{{\cal J}^-_{3}}  +  {{\cal J}^0_{33}}\,{{\cal J}^-_{2}}  - {{\cal J}^0_{12}}\,{{\cal J}^-_{3}}\,\bigg]\ +
\\ &
% \bigg[
  \  \om \,\bigg[   (4a+b+c){\cal J}^0_{11} + (4b+a+c){\cal J}^0_{22}+ (4c+a+b){\cal J}^0_{33} + (a-c){\cal J}^0_{12}+ (a-b){\cal J}^0_{13}
\\ &
  + (b-a){\cal J}^0_{23} + (b-c){\cal J}^0_{21}+ (c-a){\cal J}^0_{32} + (c-b){\cal J}^0_{31}  \bigg]\ - \ \frac{2\,d}{m}\,({{\cal J}^-_{1}}+{{\cal J}^-_{2}}+{{\cal J}^-_{3}})  \ .
%\bigg] \ .
\end{aligned}
\end{equation}

It can be shown that the primitive QES problem (\ref{Hp}) with potential (\ref{QES-privitive}) and the ground state given by (\ref{psiQES-3}) does not admit extension to a more general QES problem.

\bigskip

\subsubsection{ {\it $a=b=c$}}

Let us consider the case of three particles of equal masses and the equal spring constants: $m_1=m_2=m_3=m$ and $a=b=c$. The potentials (\ref{V3-es}), (\ref{V3-es-meq}) degenerate to a type of three-dimensional isotropic 3-body harmonic oscillator without separation of  $\rho$-variables
\begin{equation}
\label{V3-es-meq-aeq}
   V^{(3a)}\ =\ \frac{3}{2}\,m\,a^2\,\om^2\,(\rho_{12} \ + \ \rho_{13} \ + \ \rho_{23}) \ ,
\end{equation}
\cite{TME3-d}.
In this case the wavefunction (\ref{Psi03}) and the corresponding ground state energy (\ref{E0en}) take the form
\begin{equation}
\label{Psi03-3a}
   \Psi_0^{(3a)}\ =\ e^{-\frac{\om\,m}{2}\,a\,(\,\rho_{12}\ +\ \rho_{13}\ +\ \rho_{23}\,)}\ ,
\end{equation}
\begin{equation}
\label{E03-3a}
E_0\ = \ 3\,\om \,d\,a \ ,
\end{equation}
respectively. The algebraic radial Schr\"odinger operator (\ref{H-3body-algebraic-meq}) is given by
\[
 h^{(ex)}(\rho) \ = \
     -\frac{2}{m}\,\bigg[  2\, (\,\rho_{12}\, \pa_{\rho_{12}}^2\ + \
    \rho_{13}\, \pa_{\rho_{13}}^2\ +\
       \rho_{23}\, \pa_{\rho_{23}}^2\,)\ +
\]
\begin{equation}
\label{algebraic-meq-3a}
\begin{aligned}
&  {(\rho_{13} + \rho_{12} - \rho_{23})}\pa_{\rho_{13},\rho_{12}}\ +
   {(\rho_{23} + \rho_{12} - \rho_{13})}\pa_{\rho_{23},\rho_{12}}\ +
  {(\rho_{13} + \rho_{23} - \rho_{12})}\pa_{\rho_{13},\rho_{23}} \bigg]\ +
\\ &
% \bigg[
   \hskip 1cm  6\,a\,\om\,
   (\rho_{12}\pa_{\rho_{12}}\,+\,\rho_{13}\pa_{\rho_{13}}\,+\,\rho_{23}\pa_{\rho_{23}})
    \ -\ \frac{2\,d}{ m}\,\bigg(\pa_{\rho_{12}}\,+\,\pa_{\rho_{13}}\,+\,\pa_{\rho_{23}} \bigg)
    \ ,
\end{aligned}
\end{equation}
or, in its $sl(4)$-Lie-algebraic form (\ref{H-3body-Lie-algebraic-meq}),
\[
 h^{(ex)}({\cal J}) \ = \
     -\frac{2}{m}\,\bigg[  2\, ({{\cal J}^0_{11}}\,{{\cal J}^-_{1}} \ + \
       {{\cal J}^0_{22}}\,{{\cal J}^-_{2}}  \ +\
        {{\cal J}^0_{33}}\,{{\cal J}^-_{3}} ) \ + \ {{\cal J}^0_{22}}\,{{\cal J}^-_{1}}  +  {{\cal J}^0_{11}}\,{{\cal J}^-_{2}}  - {{\cal J}^0_{31}}\,{{\cal J}^-_{2}}\ +
\]
\begin{equation}
\label{Lie-algebraic-meq-3a}
\begin{aligned}
&
  {{\cal J}^0_{33}}\,{{\cal J}^-_{1}}  +  {{\cal J}^0_{11}}\,{{\cal J}^-_{3}}  - {{\cal J}^0_{23}}\,{{\cal J}^-_{1}} \ + \
 {{\cal J}^0_{22}}\,{{\cal J}^-_{3}}  +  {{\cal J}^0_{33}}\,{{\cal J}^-_{2}}  - {{\cal J}^0_{12}}\,{{\cal J}^-_{3}}\,\bigg]\ +
\\ &
% \bigg[
  6\, a\,\omega \,(   {{\cal J}^0_{11}} + {{\cal J}^0_{22}}+{{\cal J}^0_{33}} )\ -\ \frac{2\,d}{m}\,({{\cal J}^-_{1}}+{{\cal J}^-_{2}}+{{\cal J}^-_{3}})  \ .
%\bigg] \ .
\end{aligned}
\end{equation}
Now the spectrum of (\ref{algebraic-meq-3a}) is the following:
\begin{equation}
\label{ep-3b-meq-3a}
  \vep\ =\ 6\,a\,\om\,(N_1+N_2+N_3)\ ,\
\end{equation}
where $N_1,N_2,N_3=0,1,2,\ldots$ are quantum numbers.

In this case of equal masses and spring constants there exists a {\it true} $sl(4)$-QES extension, described in \cite{TME3-d}, where all eigenfunctions are proportional to
\begin{equation}
\label{psiQES-3-equal-mass}
 \tilde {\Psi}_0 \ =\ \Psi_0^{(3a)} \times e^{-(A_{12}\,\rho_{12}^2+A_{13}\,\rho_{13}^2+A_{23}\,\rho_{23}^2)} \ ,
\end{equation}
see (\ref{Psi03-3a}), thus, the exponential is a second degree polynomial in $\rho$'s.

\subsection{Atomic case: $m_1=\infty$}

An interesting special case of the three-body problem (\ref{H-3-body}) emerges when $m_1 \rar \infty$ and other two masses are kept equal $m_2=m_3=m$. In this case the potential (\ref{V3-es}) reduces to
\begin{equation}
\label{V3-es-at}
 V^{(at)}\ =\ m\,\om^2\,\bigg[   (2 \,a^2+a\, c-b \,c)\,\rho_{12} \ + \   (2 \,b^2+b\, c-a \,c)\,\rho_{13} \ + \   c\,(a+b+c)\,\rho_{23} \bigg]\ ,
\end{equation}
c.f. (\ref{V3-es-meq}).
In general, the limit $m_1 \rightarrow \infty$ when keeping $m_{2,3}$ finite corresponds to physical atomic systems where one mass is much heavier than the others (for instance, as in the negative hydrogen ion $H^- (p,e,e)$ or the helium atom He $(\al, e, e)$). We call this case {\it atomic} and for simplicity we put $m_2=m_3=m$.

For the atomic case the ground state function (\ref{Psi03}) is simplified to
\begin{equation}
\label{Psi03-at}
   \Psi_0^{(at)}\ =\ e^{-\frac{\om\, m}{2}\,(2\,a\,\rho_{12}\ + \ 2\,b\,\rho_{13}\ +\ c\,\rho_{23}\,)}\ .
\end{equation}
Since the general ground state energy (\ref{E0en}) does not depend on masses, it reads
\begin{equation}
\label{e03-at}
E_0\  =  \ \om \,d\,(a\,+\,b\,+\,c)   \ ,
\end{equation}
in this case.
The algebraic radial Schr\"odinger operator (\ref{H-3body-algebraic}) becomes
\[
 h^{(ex)}(\rho) \, = \,
     -\frac{2}{m}\,\bigg[  \rho_{12}\, \pa_{\rho_{12}}^2 +
     \rho_{13}\, \pa_{\rho_{13}}^2 +
      2\, \rho_{23}\, \pa_{\rho_{23}}^2\ +  {(\rho_{23} + \rho_{12} - \rho_{13})}\pa_{\rho_{23},\rho_{12}} +
  {(\rho_{13} + \rho_{23} - \rho_{12})}\pa_{\rho_{13},\rho_{23}}  \bigg]
\]
\begin{equation}
\label{algebraic-atomic}
\begin{aligned}
&
% \bigg[
  \qquad  + \ \om \,\bigg[\,(4a+c)\,\rho_{12}\,\pa_{\rho_{12}} + (4b+c)\,\rho_{13}\pa_{\rho_{13}}
   + 2(2c+a+b)\,\rho_{23}\pa_{\rho_{23}}
\\ &
   \hskip 0.8cm - \ c \, \left(\rho _{13}-
   \rho_{23}\right)\pa_{\rho_{12}}\ - \
     c\, \left(\rho_{12}-\rho _{23}\right)\pa_{\rho_{13}}\ + \ 2\,(a-b) \, \left(\rho _{12}-\rho _{13}\right)\pa_{\rho_{23}}\,\bigg]
\\ &
   \hskip 4cm  \ - \ \frac{d}{ m}\,\bigg[\pa_{\rho_{12}}\ +\ \pa_{\rho_{13}}\ +\ 2\,\pa_{\rho_{23}}\bigg] \ ,
\end{aligned}
\end{equation}
while in its Lie-algebraic operator (\ref{H-3body-Lie-algebraic}) form
\[
 h^{(ex)}({\cal J}) \ = \
     -\frac{2}{m}\,\bigg[   {{\cal J}^0_{11}}\,{{\cal J}^-_{1}} \ + \
       {{\cal J}^0_{22}}\,{{\cal J}^-_{2}}  \ +\
      2\,  {{\cal J}^0_{33}}\,{{\cal J}^-_{3}}  \ +  \ {{\cal J}^0_{33}}\,{{\cal J}^-_{1}} \  + \  {{\cal J}^0_{11}}\,{{\cal J}^-_{3}}
\]
\begin{equation}
\label{Lie-algebraic-atomic}
\begin{aligned}
&
-\ {{\cal J}^0_{23}}\,{{\cal J}^-_{1}}  + {{\cal J}^0_{22}}\,{{\cal J}^-_{3}}  +  {{\cal J}^0_{33}}\,{{\cal J}^-_{2}}   - {{\cal J}^0_{12}}\,{{\cal J}^-_{3}}\,\bigg]\ + \ \omega\,\bigg[ 2\,a\,( 2\,{\cal J}_{11}^0+{\cal J}_{13}^0+{\cal J}_{33}^0-{\cal J}_{23}^0)
\\ &
 + \ 2\,b\,( 2\,{\cal J}_{22}^0+{\cal J}_{23}^0+{\cal J}_{33}^0-{\cal J}_{13}^0)  + \  c\, \left( 4\,{\cal J}_{33}^0+ {{\cal J}^0_{22}}+{{\cal J}^0_{32}}-{{\cal J}^0_{12}}+{{\cal J}^0_{11}}+{{\cal J}^0_{31}}-{{\cal J}^0_{21}}\right)\,\bigg]
\\ & \ - \ \frac{d}{m}({{\cal J}^-_{1}}+{{\cal J}^-_{2}}\,+\,2 \,{{\cal J}^-_{3}})\ .
%\bigg] \ .
\end{aligned}
\end{equation}
For the atomic case $m_1\rightarrow \infty$, the co-metric defined by the coefficients in front of second derivatives in (\ref{algebraic-atomic})
\begin{equation}
\label{gmn33-rhoatomic}
 g^{\mu \nu}_{(at)}(\rho)\ = \frac{1}{m}\ \left|
 \begin{array}{ccc}
 2\, \rho_{12} & \ 0 & \ {(\rho_{23} + \rho_{12} - \rho_{13})} \\
            &                                   &                                   \\
 0 &  2\, \rho_{13} & \ {(\rho_{13} + \rho_{23} - \rho_{12})} \\
            &  \                                  &                                   \\
 {(\rho_{23} + \rho_{12} - \rho_{13})} & \ {(\rho_{13} + \rho_{23} - \rho_{12})} & 4\, \rho_{23}
 \end{array}
               \right| \ ,
\end{equation}
is proportional to $1/m$ and possesses a factorizable determinant
\[
 D_{(at)}\ \equiv \ {\rm Det} g^{\mu \nu}_{(at)}\ =\ \frac{2}{m^3} \,\left(\,\rho_{12}+\rho_{13}\,\right)\times
\]
\begin{equation}
\label{gmn33-rho-det-Matomic}
                     \left(2\,\rho_{12}\,\rho_{13} + 2\, \rho_{12}\,\rho_{23} + 2\, \rho_{13}\,\rho_{23}-\rho_{12}^2- \rho_{13}^2 - \rho_{23}^2\right)\ =
                     \frac{2}{m^3} \,\left(\,\rho_{12}+\rho_{13}\,\right)\,S^2_{\triangle}
                      \ ,
\end{equation}
which is positive definite (cf. (\ref{gmn33-rho-det-M})). We emphasize that the operator (\ref{algebraic-atomic}) is three-dimensional, all three $\rho$-variables remain dynamical, see discussion Section III.C.

It can be shown that the primitive QES problem (\ref{Hp}) with potential (\ref{QES-privitive}) and  ground state given by (\ref{psiQES-3}) does not admit extension to a more general QES problem as in the case of equal masses but non-equal spring constants.

\subsection{Two-center case, $m_2,m_3 = \infty$}

In the genuine two-center case two masses are considered infinitely heavy, $m_{2,3} \rar \infty$, thus, the reduced mass $\mu_{23}$ also tends to infinity, while the third mass $m_1=m$ remains finite. Sometimes it is called the Born-Oppenheimer approximation of zero order. It implies that the coordinate $\rho_{23}$ is classical (thus, unchanged in dynamics, being constant of motion in the process of evolution, it can be treated as external parameter), while two other $\rho$-variables remain dynamical. The 3-body problem is converted to a two-center problem. The potential (\ref{V3-es}) depends on masses, and in order to keep it finite in the molecular limit, we set the spring constant $c=0$ from the very beginning. The ground state energy does not depend on masses and is equal to
\begin{equation}
\label{E0-c}
     E_0^{(c=0)}\  =  \ \om \,d\,(a+b)\ ,
\end{equation}
c.f. (\ref{E0en}). The energy is measured from the minimum of the potential (\ref{V3-es}), $V_{min}=V(0)=0$.

In the limit $m_{2,3} \rar \infty$ the radial Laplacian (\ref{addition3-3r-M}) (as well as the associated Laplace-Beltrami operator) loses the property of invert ability: both 3rd row and 3rd column in (\ref{gmn33-rho}) vanish as well as the determinant of co-metric $D$ (\ref{gmn33-rho-det-M}). However, the radial Schr\"odinger operator with potential (\ref{V3-es}) at $c=0$ in the molecular limit is well-defined and finite. In particular,
\begin{equation}
\label{mol-case}
  \frac{1}{2}\De^{(mol)}_{\rm rad}(\rho)\ =\
  \frac{1}{m} \left(\rho_{12}\, \pa_{\rho_{12}}^2 +
   \rho_{13}\, \pa_{\rho_{13}}^2 +
  (\rho_{13} + \rho_{12} - \rho_{23})\pa_{\rho_{13},\,\rho_{12}}\ + \
   \frac{d}{2}\,(\pa_{\rho_{12}} + \pa_{\rho_{13}}) \right) \ ,
\end{equation}
where now $\rho_{23}$ plays the role of a  parameter. It must be noted that the {\it same} expression is obtained directly from (\ref{Tflat}), thus before center-of-mass separation,
by taking the limit $m_{2,3} \rar \infty$, then rewriting the first term $\De_1^{(d)}$ in
new variables
\[
   ({\bf r_1})\ =\ (\rho_{12}=r_{12}^2,\ \rho_{13}=r_{13}^2,\ \Om_1)\ ,
\]
with $\rho_{23}=r_{23}^2$ kept fixed and separating out the $(d-2)$ angular variables $\{\Om_1\}$.

\begin{figure}
  \centering
  \includegraphics[width=12.0cm]{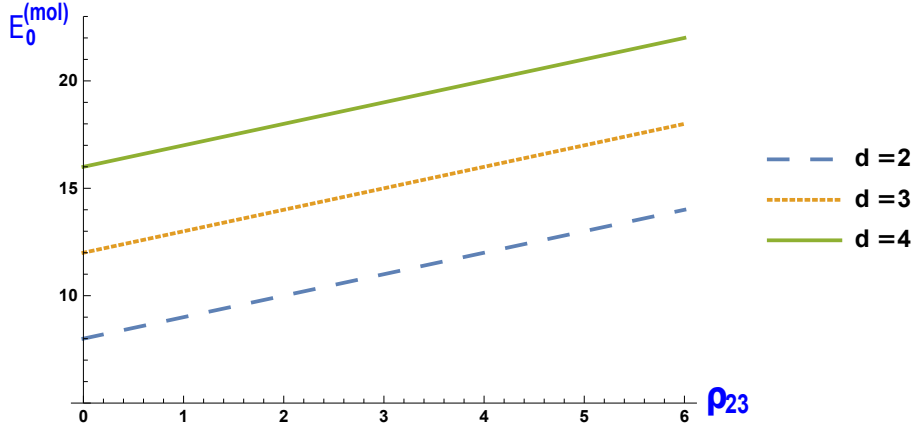}
  \caption{Ground state energy potential curves (\ref{E0-mol}) {\it vs} $\rho_{23}$ (classical coordinate) at $d=2,3,4$. It corresponds to the values of the parameters $a=b=m=\om=1\,$ .  }
  \label{3Bm}
\end{figure}

Hence, in the molecular case the spectral problem (\ref{H-3-body}) becomes two-dimensional and the potential (\ref{V3-es}) at $c=0$ simplifies to
\begin{equation}
\label{V-mol-case}
   V^{(mol)}\ =\ 2\,m\,\om^2\,(a+b)\,\bigg[a\,\rho_{12} \ + \ b\,\rho_{13} \bigg]\ ,
\end{equation}
hence, $\nu_{23}=0$ \footnote{
We must emphasize that, in general, the potential (\ref{V-mol-case}) is defined up to additive constant, which can depend on classical coordinate $\rho_{23}$. This constant defines the reference point for energy.}. The ground state function for the molecular Hamiltonian (\ref{mol-case}) + (\ref{V-mol-case})
\begin{equation}
\label{H-mol}
   {\cal H}^{(mol)}\ =\ -\De^{(mol)}_{\rm rad}(\rho)\ +\ V^{(mol)}\ ,
\end{equation}
can be easily found
\begin{equation}
\label{Psi03-mol}
   \Psi_0^{(mol)}\ =\ e^{-\om\,m\,(\,a\,\rho_{12}\ +\ b\,\rho_{13}\,)}\ ;
\end{equation}
it corresponds to the energy
\begin{equation}
\label{E0-mol}
E_0^{(mol)}\  =  \ \om \,d\,(a+b)\ +\ 2\,m\,\om^2 \,a\,b\,\rho_{23}   \ ,
\end{equation}
c.f. (\ref{E0-c}). Note that $E_0^{(mol)}$ is measured from the reference point $V^{(mol)}(0)=0$ and it is always larger than (or equal to) (\ref{E0-c}), $E_0^{(mol)} \geq E_0^{(c=0)}$.
$E_0^{(mol)}$ takes minimal value at $\rho_{23}=0$, where it coincides with exact ground state energy (\ref{E0-c}).
From the viewpoint of the  Born-Oppenheimer approximation, widely used in molecular physics, where the first particle $m_1=m$ can be associated with an electron and the whole system with a one-electron homonuclear diatomic ion, the Hamiltonian ${\cal H}^{(mol)}$ (\ref{H-mol})
%\begin{equation}
%\label{H-mol}
%   {\cal H}^{(mol)}\ =\ -\frac{1}{2}\De^{(mol)}_{\rm rad}(\rho)\ +\ V^{(mol)}\ ,
%\end{equation}
has the meaning of the so-called electronic Hamiltonian. It describes the electronic
degrees of freedom of a molecular system. The ground state energy $E_0^{(mol)}$ (\ref{E0-mol})
is called the {\it ground state energy potential curve} or, simply speaking, the potential curve, see Fig. \ref{3Bm}. The Hamiltonian ${\cal H}^{(mol)}$ (\ref{H-mol}) describes two-center problem.

From the coefficients in front of second derivatives in the operator $\frac{1}{2}\De^{(mol)}_{\rm rad}$ (\ref{mol-case}) one can  form the matrix
\begin{equation}
\label{gmn-mol}
g_{(mol)}^{\mu \nu} \ = \ \frac{1}{m}\left(
\begin{array}{cc}
 \rho_{12} & \frac{1}{2} (\rho_{12}+\rho_{13}-\rho_{23}) \\
 \frac{1}{2} (\rho_{12}+\rho_{13}-\rho_{23}) & \rho_{13} \\
\end{array}
\right) \ ,
\end{equation}
with determinant
\begin{equation}
\label{det-mol}
   D_{(mol)} \ \equiv \ \text{Det}(g_{(mol)}^{\mu \nu}) \ = \ \frac{1}{4\,m^2}\big[ 2\,\rho_{23}\,(\rho_{12}+\rho_{13}) -\ (\rho_{12}-\rho_{13})^2 \  \ - \ \rho_{23}^2\,\big]
   \ =\ \frac{1}{4\,m^2}\,S^2_{\triangle}\ ,
\end{equation}
cf. (\ref{gmn33-rho-det-M}), (\ref{gmn33-rho-det-Matomic}), which remains positive definite. The matrix (\ref{gmn-mol}) is the principal minor for $g^{33}$.

Making a gauge rotation of the operator $\frac{1}{2}\De^{(mol)}_{\rm rad}$ with the gauge factor
\[
       \Gamma \ = \ D_{(mol)}^{\frac{2-d}{4}} \ \ ,
\]
we obtain a Laplace-Beltrami operator with co-metric $g_{(mol)}^{\mu \nu}$ plus effective potential,
\[
  \Gamma^{-1}\, \frac{1}{2}\De^{(mol)}_{\rm rad} \,\Gamma \ = \
  \De^{(mol)}_{LB} - V^{(mol)}_{\text{eff}} \ .
\]
Here,
\[
 V^{(mol)}_{\text{eff}} \ = \ \frac{(d-2)(d-4)\,\rho_{23}}{16\,m^2\,D_{(mol)}} \ ,
\]
is proportional to the classical coordinate $\rho_{23}$ and vanishes at $d=2,4$.
Hence, the matrix $g_{(mol)}^{\mu \nu}$ has the  meaning of a co-metric.

The radial Schr\"odinger operator (\ref{H-3body-algebraic}) in the molecular limit remains algebraic:
\begin{equation}
\label{algebraic-molecular}
\begin{aligned}
 h^{(ex)}(\rho) \ & = \
     -\frac{2}{m}\, \bigg[  \rho_{12}\, \pa_{\rho_{12}}^2 +
     \rho_{13}\, \pa_{\rho_{13}}^2 \ +  {(\rho_{13} + \rho_{12} - \rho_{23})}\pa_{\rho_{13},\rho_{12}} \bigg] \ +
\\ &\ 2\, \om  \left[\, (2\,a+b)\,\rho_{12}\,\pa_{\rho_{12}}  \, + \, (2\,b+a)\,\rho_{13}\,\pa_{\rho_{13}} \,+\,b\,(\rho_{13}-\rho_{23})\,\pa_{\rho_{12}} \,+\,a\,(\rho_{12}-\rho_{23})\,\pa_{\rho_{13}}  \,\right]
\\ &  \, - \ \frac{d}{m}\,(\pa_{\rho_{12}} \ +\ \pa_{\rho_{13}}) \ .
%\bigg] \ .
\end{aligned}
\end{equation}
Its eigenfunctions $\phi_{k_1,k_2}$ are marked by two integer (quantum) numbers $k_1,k_2=0,1,\ldots$. They are two-variate triangular polynomials in $\rho_{12}, \rho_{13}$ and its spectrum $\varepsilon_{k_1,k_2}$ is linear in quantum numbers $(k_1,k_2)$ and  $\varepsilon_{0,0}=0$. The spectrum of the Hamiltonian (\ref{H-mol}), is
\begin{equation}
\label{Ek-mol}
  E^{(mol)}_{k_1,k_2} = E_0^{(mol)} + \varepsilon_{k_1,k_2} \ .
\end{equation}
Note that the operator ${\cal H}^{(mol)}$ (\ref{H-mol}) is self-adjoint with respect to the measure
\[
 w\ \equiv \  \Gamma^2\,\sqrt{-D_{(mol)}}\ \sim \ D_{(mol)}^{\frac{3}{2} - 2\,d} \ .
\]
The algebraic operator (\ref{H-3body-Lie-algebraic}) is also Lie-algebraic of the form
\[
 h^{(ex)}({\cal J}) \ = \
     -\frac{2}{m}\,\bigg[  {{\cal J}^0_{11}}\,{{\cal J}^-_{1}} \ + \
     {{\cal J}^0_{22}}\,{{\cal J}^-_{2}}   \ + \ {{\cal J}^0_{22}}\,{{\cal J}^-_{1}}  +  {{\cal J}^0_{11}}\,{{\cal J}^-_{2}}  - \rho_{23}{{\cal J}^-_{1}}\,{{\cal J}^-_{2}}\,\bigg]\ +
\]
\begin{equation}
\label{}
\begin{aligned}
% \bigg[
&  \ 2\,\om\,\left[ \,(2\,a+b){{\cal J}^0_{11}}+(2\,b+a){{\cal J}^0_{22}} \ +\ \,b\,({{\cal J}^0_{21}}-\rho_{23}{{\cal J}^-_{1}})+a\,({{\cal J}^0_{12}}-\rho_{23}{{\cal J}^1_{2}})\right]  \ - \ \frac{d}{m}({{\cal J}^-_{1}}+{{\cal J}^-_{2}})\ ,
%\bigg] \ .
\end{aligned}
\end{equation}
where the  $J$'s are generators of the algebra $b_3 \in sl(3,{\bf R})$ realized by the first order differential operators.
Hence, the hidden algebra in the molecular limit is $b_3 \in sl(3,{\bf R})$, contrary to the general case and all other particular cases, when the hidden algebra is $b_4 \in sl(4,{\bf R})$. In the present case $\rho_{23}$ plays the role of a parameter. Note that the algebra $b_3$ is six-dimensional (of lower triangular matrices of size $3 \times 3$), it is spanned by $\{ {\cal J}^0_{11}, {\cal J}^0_{12}, {\cal J}^0_{21}, {\cal J}^0_{22}, {\cal J}^-_{1}, {\cal J}^-_{2}  \}$.

It can be shown that the primitive QES problem (\ref{Hp}) with potential (\ref{QES-privitive}) and  ground state given by (\ref{psiQES-3}) at $c=0$, see (\ref{Psi03-mol}) does not admit extension to a more general QES problem.

\bigskip

\subsection{Molecular case in the Born-Oppenheimer approximation: $m_2,m_3 \gg m_1$}

\bigskip

For the molecular case, $m_1 \ll m_2,m_3 < \infty$, the 3-body oscillator model we deal with allows exact solvability
and a critical analysis of the Born-Oppenheimer approximation as described below.

Following the formalism of the Born-Oppenheimer approximation, the energy of $E^{(mol)}_{k_1,k_2}$ (\ref{Ek-mol}) of the electronic Hamiltonian $ {\cal H}^{(mol)}$ (\ref{H-mol}) should appear as the potential in a two-body {\it nuclear} Hamiltonian.
In order to derive this Hamiltonian we should replace in (\ref{Hgen}) the first term of kinetic energy by (\ref{Ek-mol}), separate center-of-mass in the second+third terms and introduce the Euler coordinates
\[
   ({\bf r}_2\ ,\ {\bf r}_3) \rar ({\bf R}_{0}\ ,\ \rho=r_{23}^2\ ,\ \Om_{23})\ ,
\]
and then separate out $(d-1)$ angular variables $\{ \Om_{23} \}$. As a result
we arrive at
\begin{equation}
\label{Hnucl}
 {\cal H}^{(nucl)}\ =\ -\frac{1}{\mu}
    \bigg(2 \,\rho\, \pa^2_{\rho}\ +\ d\,\pa_{\rho} \bigg)\ +\ \frac{L (L+d-2)}{\rho}\ +\  2\,\om^2(\,m \,a\,b+\nu_{23})\,\rho
     + \,\om \,d\,(a+b)\ +\ \varepsilon_{k_1,k_2}\ ,
\end{equation}
where $\mu \equiv \mu_{23} = \frac{m_2 m_3}{m_2+m_3}$ is the reduced mass, $m_{2,3}$ are masses of the nuclei, $\nu_{23}$ given by (\ref{freq-3}) and $L$ is its two-body nuclear angular momentum and $m=m_1$. Now $\rho \equiv \rho_{23}$ is restored as a dynamical variable.
For simplicity we limit ourselves to the ground state, putting $L=0$ and $k_1=k_2=0$ in the Hamiltonian (\ref{Hnucl}),
\begin{equation}
\label{Hnucl-0}
 {\cal H}_0^{(nucl)}\ =\ -\frac{1}{\mu}
    \bigg(2 \,\rho\, \pa^2_{\rho}\ +\ d\,\pa_{\rho} \bigg)\ +\  2\,\om^2(\,m \,a\,b+\nu_{23})\,\rho
   \  + \ \om \,d\,(a+b)\ ,
\end{equation}
c.f. (\ref{H-2-body}) with potential (\ref{V2-exact}).
This nuclear Hamiltonian defines the so-called vibrational spectrum of the ground state, its lowest eigenvalue (sometimes called zero-point energy) is
\begin{equation}
\label{Enucl-0}
    E_0^{(nucl)}\ =\ \left(\om\,d\,( a+ b) \, + \,\om\,d\, \sqrt{\frac{a\,b\,m}{\mu}\bigg(1 \,+\, \frac{\nu_{23}}{a\,b\,m}\bigg)} \,\right)\ .
\end{equation}
Making comparison of the exact energy $E_0$ (\ref{E0en}) with $E_0^{(nucl)}$ we get
\begin{equation}
\label{del-Enucl-1}
   E_0^{(nucl)}\ -\ E_0 \ = \ \om\,d\,\bigg\{ \sqrt{\frac{a\,b\,m}{\mu}\bigg(1 \,+\, \frac{\nu_{23}}{a\,b\,m}\bigg)} \ - \ c\bigg\} \ .
\end{equation}
This difference ``measures" the accuracy of the Born-Oppenheimer approximation: it tends to zero as $\mu \rar \infty$. For the relevant physics case $m_2=m_3=1$, we obtain the following expansion in powers of the small parameter $m \equiv m_1 \ll 1$
\begin{equation}
\label{del-Enucl-2}
   E_0^{(nucl)}\ -\ E_0 \ = \ \frac{\omega\,d}{2}\,\bigg(    (a+b)\,m \ - \ \frac{ \left(a^2-14 a b+4 a c+b^2+4 b c\right)}{4\, c}\,m^2 \ + \ \ldots    \bigg) \ .
\end{equation}

As previously observed in the one-dimensional case $d=1$ \cite{Fernandez}, the Born-Oppenheimer approximation yields the leading term of the expansion of the exact result in powers of $m$ (or the ratio of the electron to nuclear mass) for any $d>1$.

\subsection{One-dimensional case $d=1$}

For $d=1$ (three particles on the line), the 3-body system is described by the
Hamiltonian
\begin{equation}
\label{H1D}
  H \ = \ -\bigg( \,\frac{1}{2\,m_1}\pa^2_{x_1} \ + \ \frac{1}{2\,m_2}\pa^2_{x_2} \ + \ \frac{1}{2\,m_3}\pa^2_{x_3}  \, \bigg) \ + \ V(x_{12},\,x_{23},\,x_{31}) \ ,
\end{equation}
with the potential $V$ which depends on relative distances
\[
x_{12} = |x_1-x_2|\ ,\ x_{13} = |x_1-x_3| \ ,\ x_{23} = |x_2-x_3|\ ,
\]
where only two of them are independent.
One can separate out the center-of-mass variable (\ref{CMS}); then assuming that $x_{23}$ is a dependent variable, $x_{23} = x_{13} - x_{12}$\footnote{It corresponds to particle ordering
$x_1\leq x_2 \leq x_3$}, we arrive at the two-dimensional spectral problem for the radial Hamiltonian
\begin{equation}
\label{H1Drad}
  H_{rad} \ = \ -\frac{1}{2\,\mu_{12}}\pa^2_{x_{12}} \ - \ \frac{1}{2\,\mu_{13}}\pa^2_{x_{13}} \ - \ \frac{1}{m_1}\pa^2_{x_{12},x_{13}} + \ V(x_{12},\,x_{13})      \ .
\end{equation}
see e.g. \cite{TME3-d}. For the case of a 3-body harmonic oscillator, the potential is given by (\ref{V3-es}) at $d=1$,
\[
   V_{d=1}^{(es)}\ =\ 2\,\om^2\bigg[\nu_{12}\,x_{12}^2 \ + \ \nu_{13}\,x_{13}^2 \ +
   \ \nu_{23}\,x^2_{23} \bigg]\ ,
\]
with $\nu$'s from (\ref{freq-3}) and $x^2_{23}= (x_{13} - x_{12})^2$, see \cite{Fernandez}. Its final form is
\begin{equation}
\label{V3-d=1}
  V_{d=1}^{(es)}\ =\ 2\,\om^2\bigg[(\nu_{12}+\nu_{23})\,x_{12}^2 \ + \ (\nu_{13}+\nu_{23})\,x_{13}^2 \ -\ 2\,\nu_{23}\,x_{12}\,x_{13} \bigg]\ .
\end{equation}

It corresponds to the radial Hamiltonian
\begin{equation}
\label{H3-d1}
    H_{d=1} \ = \ -\frac{1}{2\,\mu_{12}}\pa^2_{x_{12}} - \frac{1}{2\,\mu_{13}}\pa^2_{x_{13}} - \frac{1}{m_1}\pa^2_{x_{12},x_{13}} + \ 2\,\om^2\bigg[(\nu_{12}+\nu_{23})\,x_{12}^2 \ + \ (\nu_{13}+\nu_{23})\,x_{13}^2 \ -\ 2\,\nu_{23}\,x_{12}\,x_{13} \bigg]      \ .
\end{equation}
In is easy to check that the ground state function in (\ref{H3-d1}) is given by
\begin{equation}
\label{Psi03d1}
   \Psi_0^{(es)}\ =\ e^{-\om\, (a\,\mu_{12}\,x^2_{12}\,+\,b\,\mu_{13}\,x^2_{13}\,+\,c\,\mu_{23}\,(x_{13} - x_{12})^2)}\ ,
\end{equation}
with the ground state energy
\begin{equation}
\label{E0end1}
E_0\ = \ \omega \, (\,a\ +\ b\ +\ c \,) \ .
\end{equation}
By making a gauge rotation with $\Psi_0^{(es)}$ (\ref{Psi03d1}) as the gauge factor, the potential in the radial Schr\"odinger operator $H_{d=1}$ (\ref{H3-d1}) disappears and we arrive at the algebraic operator with polynomial coefficients
\[
 h^{(es)} \ \equiv\  \big( \Psi_0^{(es)}\big)^{-1}\, [\,H_{d=1}\,- \,E_0\,]\, \Psi_0^{(es)}
 \ =
\]
\[
    -\,\frac{1}{2\,\mu_{12}}\pa^2_{x_{12}}\ -\ \frac{1}{2\,\mu_{13}}\pa^2_{x_{13}}\ - \ \frac{1}{m_1}\pa^2_{x_{12},x_{13}} \ +
\]
\begin{equation}
\label{H-3body-algebraic1D}
\begin{aligned}
&  \frac{2 \,\omega\,  \left[\,\mu _{12} \left(a\, m_1\, x_{12}\,+\,b\, \mu _{13}\, x_{13}\right)\ + \ c\, \mu _{23}\, \left(x_{12}-x_{13}\right) \left(m_1-\mu _{12}\,\right)\right]}{\mu _{12}\, m_1}\,\pa_{x_{12}} \ + \\ &
\frac{2 \,\omega\,  \left[\,\mu _{13} \,\left(a \,\mu _{12}\, x_{12}\,+\,b\, m_1\, x_{13}\,\right)\ + \ c\, \mu _{23}\, \left(x_{13}-x_{12}\right) \left(m_1-\mu _{13}\,\right)\right]}{\mu _{13}\, m_1}\,\pa_{x_{13}} \ .
\end{aligned}
\end{equation}
cf. \cite{Ruhl:1995}.
Here $E_0$ is given by (\ref{E0end1}). This operator has a Lie-algebraic form: it can be rewritten in terms of the generators of the algebra $b_3 \in sl(3, {\bf R})$ (see e.g. \cite{Turbiner:1988,Turbiner:2016})
\begin{eqnarray}
\label{sl3R}
 {\cal J}_i^- \ &=& \ \frac{\pa}{\pa u_i}\ ,\qquad \quad i=1,2\ , \non  \\
 {{\cal J}^0_{ij}} \ &=& \
               u_i \frac{\pa}{\pa u_j}\ , \qquad i,j=1,2\ ,\non \\
 {\cal J}^0(N)\  &=& \ \sum_{i=1}^{2} u_i \frac{\pa}{\pa u_i}-N\, , \non \\
 {\cal J}_i^+(N) \ &=& \ u_i \,{\cal J}^0(N)\ =\
    u_i\, \left( \sum_{j=1}^{2} u_j\frac{\pa}{\pa u_j}-N \right)\ ,
       \quad i=1,2\ ,
\end{eqnarray}
where $N$ is a parameter; it is denoted
\[
 u_1\ \equiv \ x_{12}\ ,\qquad u_2\ \equiv \ x_{13} \ .
\]
Explicitly,
\[
 h^{(es)} \ = \   -\,\frac{1}{2\,\mu_{12}}\,{({\cal J}_1^-)}^2 \ -\ \frac{1}{2\,\mu_{13}}\,{({\cal J}_2^-)}^2 \ - \ \frac{1}{m_1}\,{\cal J}_1^-\,{\cal J}_2^- \ +
\]
\begin{equation}
\label{H-3body-algebraic1Dslr}
\begin{aligned}
&  \frac{2 \,\om\,  \left[\,\mu_{12} \left(a\, m_1\, {{\cal J}^0_{11}}\,+\,b\, \mu_{13}\,
  {{\cal J}^0_{21}}\right)\ + \ c\, \mu_{23}\, \left({{\cal J}^0_{11}}-{{\cal J}^0_{21}}\right) \left(m_1-\mu_{12}\,\right)\right]}{\mu_{12}\, m_1} \ + \\
&
  \frac{2 \,\om\,  \left[\,\mu _{13} \,\left(a \,\mu_{12}\, {{\cal J}^0_{12}}\,+\,b\, m_1\, {{\cal J}^0_{22}}\,\right)\ + \ c\, \mu_{23}\, \left({{\cal J}^0_{22}}-{{\cal J}^0_{12}}\right) \left(m_1-\mu_{13}\,\right)\right]}{\mu_{13}\, m_1} \ .
\end{aligned}
\end{equation}
cf. \cite{Ruhl:1995}.

The algebraic operator (\ref{H-3body-algebraic1Dslr}) does not admit extension to a non-trivial (non primitive) QES problem.

\section{Integrability analysis of the three-body problem: arbitrary mass case}

Here we present the 1st and 2nd order integrals (symmetries) of the 3-body Hamiltonian in $S$-state for the case of arbitrary masses. We begin with
% To save space we give the details only for the classical case. The results for the quantum
% system are virtually identical.
the classical kinetic energy
\begin{equation}
\label{S1}
   \frac{1}{2}\De_{\rm rad}^{(\text{classical})}(\rho) \ \equiv \ S_1 \ = \  \frac{1}{\mu_{12}}\rho_{12}\, p_1^2 \ + \ \frac{1}{\mu_{13}}\rho_{13}\, p_2^2\ + \ \frac{1}{\mu_{23}}\rho_{23}\, p_3^2
\end{equation}
\[
   \ + \ \frac{(\rho_{12}+\rho_{13}-\rho_{23})}{m_1}p_1\, p_2\ + \ \frac{(\rho_{12}+\rho_{23}-\rho_{13})}{m_2}p_1\, p_3\ + \ \frac{(\rho_{23}+\rho_{13}-\rho_{12})}{m_3}p_2\,p_3 \ ,
\]
c.f. (\ref{addition3-3r-M}). Here $p_1=p_{\rho_{12}},p_2=p_{\rho_{13}},p_3=p_{\rho_{23}}$ are the conjugate canonical momenta. The function $\De_{\rm rad}^{(\text{classical})}$ is invariant under the ${\mathcal{S}}_3-$group action (permutation between any pair of particles).
There exists a single, 1st order in the momenta and $\rho$'s, constant of the motion,
\[
  L_0 \ = \  m_3\,[\,(m_1+m_2)\rho_{13}\ -\ (m_1+m_2)\rho_{23}\ +\ (m_1-m_2)\rho_{12}\, ]\,p_1\ +
\]
\[
  m_2\,[\,(m_1+m_3)\rho_{23}\ -\ (m_1+m_3)\rho_{12}\ +\ (m_3-m_1)\rho_{13}\,]\,p_2\ +
\]
\[
   m_1\,[\,(m_2+m_3)\rho_{12}\ -\ (m_2+m_3)\rho_{13}\ +\ (m_2-m_3)\rho_{23}\,]\,p_3\ ,
\]
whose Poisson bracket with $S_1$ vanishes: $\{S_1,\,L_0\}_{ {PB}}=0$. It is easy to check that the $L_0$ is anti-invariant under the $\mathcal{S}_3-$permutation group action. The existence of $L_0$ allows us to separate out one variable in the free Hamiltonian (\ref{S1}). This was carried out in \cite{TME3-d} where the ``ignorable" coordinate $W_3$ was separated.

There are three 2nd order integrals that are quadratic in momenta and linear in the $\rho$ coordinates: $S_1,S_2,S_3$  with vanishing Poisson brackets $\{S_j,S_k\}=0$, $1\le j,k\le 3$\ ,
\[
  S_2\ = \  \, \rho_{13}\,p_2^2 \ - \ \rho_{12}\,p_1^2  \ + \ (\rho_{23}+\rho_{13}-\rho_{12})\,p_2\,p_3\ + \ (\rho_{13}-\rho_{12}-\rho_{23})\,p_1\,p_3 \ ,
\]
\[
  S_3 \ = \  \, -\rho_{13}\,p_2^2 \ - \ \rho_{12}\,p_1^2  \ + \ (\rho_{23}-\rho_{13}-\rho_{12})\,p_1\,p_2 \ ,
\]
see (\ref{S1}) as for $S_1$. Thus, the original 3-body free system for $S$-states is integrable. Note that $S_2$ is anti-invariant under the $\mathcal{S}_2-$permutation group action (permutation between the particle 2 and particle 3) only.
Besides that there are three 2nd order integrals, those are quadratic in the $\rho$ coordinates,
\[
  F_1 \ = \ [\,\rho_{12}^2+\rho_{13}^2+\rho_{23}^2-2\,(\rho_{12}\,\rho_{13}+
  \rho_{12}\,\rho_{23}+\rho_{13}\,\rho_{23})\,](m_2\,p_2-m_3\,p_1)^2 \ ,
\]
\[
   F_2 \ = \ [\,\rho_{12}^2+\rho_{13}^2+\rho_{23}^2-2\,(\rho_{12}\,\rho_{13}+\rho_{12}\,\rho_{23}+
   \rho_{13}\,\rho_{23})\,](m_1\,p_3-m_3\,p_1)^2 \ ,
\]
\[
   F_3 \ = \ [\,\rho_{12}^2+\rho_{13}^2+\rho_{23}^2-2\,(\rho_{12}\,\rho_{13}+\rho_{12}\,\rho_{23}+
   \rho_{13}\,\rho_{23})\,](m_1\,p_3-m_2\,p_2)^2 \ ,
\]
as well as $L_0^2$. Therefore, the original classical free 3-body system for $S$-states, described by $S_1$ (\ref{S1}), is maximally superintegrable. Only 5 integrals can be functionally independent in this case, but we have not computed the dependence relations explicitly. Note that among the known integrals there are also three triplets that are in involution:
\begin{equation}
\label{trip}
   \{S_1,\,S_2,\,S_3\}\ ,\quad \{S_1,\,F_1,\,S_3\}\ ,
\end{equation}
\[
\{\,S_1,\,L_0^2,\, [m_1^2+m_1(m_2+m_3)-m_2\,m_3]\,F_1+[m_2^2+m_2(m_1+m_3)-m_1\,m_3]\,F_2
\]
\[ + \ [m_3^2+m_3(m_1+m_2)-m_1\,m_2]\,F_3 \,\}\ ,
\]
thus, forming commutative Poisson algebras. The maximal number of integrals in involution is equal to three. In general,
the Poisson bracket between two elements of different triplets is non zero. This ends our analysis of integrals of the classical 3-body free Hamiltonian.

If we take the classical 3-body harmonic oscillator by adding to $2 S_1$ the 3-body oscillator interaction potential
\begin{equation}
\label{HclaS}
  {\cal H}^{(cl)} \  \equiv  \  2\,S_1 \ + \ 2\,\om^2\,\left(\nu_{12}\,\rho_{12}\ + \ \nu_{13}\,\rho_{13}\ + \ \nu_{23}\,\rho_{23}\right)\ ,
\end{equation}
in general, none of the above-mentioned integrals can be augmented to integrals of this new system if the  $\nu$'s are arbitrary. However, in the special case
\begin{equation}
\label{super-int}
   m_2\,\nu_{13}=m_3\,\nu_{12}\ ,\qquad m_1\,\nu_{23}=m_2\,\nu_{13}\ ,\qquad m_3\,\nu_{12}=m_1\,\nu_{23}\ ,
\end{equation}
(any two above relations imply that the third relation should hold), the $L_0$ appears as an integral: $\{{\cal H}^{(cl)},\,L_0\}_{ {PB}}=0$, as do the prolonged $S_{2,3}$:
\[
 {\tilde S}_2 \equiv S_2 \ +\ \om^2\,\frac{\nu_{13}}{m_3(m_1+m_2+m_3)}\ \times
\]
\[
   \bigg(m_3(m_2^2+m_1m_3+m_2m_3)\,\rho_{13} - m_2(m_3^2+m_1m_2+m_2m_3)\,\rho_{12} - m_2\,m_3(m_2-m_3)\,\rho_{23}\,  \bigg)   \ ,
\]
\[
   {\tilde S}_3 \equiv S_3 \ + \ \om^2\,\frac{m_1\nu_{13}}{m_3(m_1+m_2+m_3)} \bigg(m_2\,m_3\,\rho_{23} - m_2(m_2+m_3)\,\rho_{12} -m_3(m_2+m_3)\,\rho_{13} \bigg)   \ .
\]
Furthermore, it can be checked that $F_1,F_2,F_3$, see above, remain integrals as well. Thus, the classical system ${\cal H}^{(cl)}$ under conditions (\ref{super-int}) is {\it maximally} superintegrable. Only 5 integrals can be functionally independent, hence, there must exist two relations between them. Note that if only the condition
\begin{equation}
\label{super-int-min}
      m_2\,\nu_{13}=m_3\,\nu_{12}\ ,
\end{equation}
holds, then it can be shown that $(\tilde S_2, F_2, F_3)$ are not conserved anymore, the classical system described by ${\cal H}^{\text{(cl)}}$ is {\it minimally} superintegrable: the triplet $(\tilde S_3, F_1, L_0)$  commutes only with the Hamiltonian.

As for the quantum 3-body harmonic oscillator, let us consider first its kinetic energy - the radial operator $\big(-\frac{\De_{\rm rad}}{2}\big)$ (\ref{addition3-3r-M}) - which is, in fact, the free 3-body Hamiltonian $S_1^{(q)}$ - the quantum counterpart of $S_1$ (\ref{S1}).
It can be shown that the quantum counterparts of $S_{2,3}$ and $F_{1,2,3}$, see above,
\begin{equation}\label{}
\begin{aligned}
S_2^{(q)} \ = & \ \rho_{13}\,\partial^2_{\rho_{13}}   \ - \ \rho_{12}\,\partial^2_{\rho_{12}} \ + \ ( \, \rho_{23} + \rho_{13} - \rho_{12}\,)\,\partial^2_{\rho_{23},\rho_{13}} \ + \ (\, \rho_{13} - \rho_{12}  - \rho_{23}\,)\,\partial^2_{\rho_{23},\rho_{12}} \\ & \ + \ \frac{d}{2}\,(\, \partial_{\rho_{13}} \ - \  \partial_{\rho_{12}}  \,) \ , \nonumber
\end{aligned}
\end{equation}
\begin{equation}\label{}
\begin{aligned}
S_3^{(q)} \ = & \ -\rho_{13}\,\partial^2_{\rho_{13}}   \ - \ \rho_{12}\,\partial^2_{\rho_{12}} \ + \ ( \, \rho_{23} - \rho_{13} - \rho_{12}\,)\,\partial^2_{\rho_{12},\rho_{13}} \ - \ \frac{d}{2}\,(\, \partial_{\rho_{13}} \ + \  \partial_{\rho_{12}}  \,) \ , \nonumber
\end{aligned}
\end{equation}
\begin{equation}\label{}
\begin{aligned}
F_1^{(q)} \ = & \  (\rho _{12}^2+\rho _{13}^2+\rho _{23}^2-2 \left(\rho _{12} \rho _{13}+\rho _{23} \rho _{13}+\rho _{12} \rho _{23}\right))(\,m_2^2\partial^2_{\rho_{13}}-2m_2m_3\partial^2_{\rho_{12},\rho_{13}}+m_3^2\partial^2_{\rho_{12}} \,)
\\ &
\ + \ (d-1)\bigg[   m_3^2  \left(\rho _{12}-\rho _{13}-\rho _{23} \right)\partial_{\rho_{12}} \ + \  m_2^2 \left(\rho _{13}-\rho _{12}-\rho _{23}\right) \partial_{\rho_{13}}
\\ &
+ \,m_3\, m_2\, (\, \left(\rho _{13}+\rho _{23}-\rho _{12}\right) \partial_{\rho_{13}} + \left(\rho _{12}-\rho _{13}+\rho _{23}\right) \partial_{\rho_{12}} \,)  \bigg]
 \ , \nonumber
\end{aligned}
\end{equation}
\begin{equation}\label{}
\begin{aligned}
F_2^{(q)} \ = & \  (\rho _{12}^2+\rho _{13}^2+\rho _{23}^2-2 \left(\rho _{12} \rho _{13}+\rho _{23} \rho _{13}+\rho _{12} \rho _{23}\right))(\,m_1^2\partial^2_{\rho_{23}}-2m_1m_3\partial^2_{\rho_{12},\rho_{23}}+m_3^2\partial^2_{\rho_{12}} \,)
\\ &
\ + \ (d-1)\bigg[   m_3^2  \left(\rho _{12}-\rho _{13}-\rho _{23} \right)\partial_{\rho_{12}} \ + \  m_1^2 \left(\rho _{23}-\rho _{12}-\rho _{13}\right) \partial_{\rho_{23}}
\\ &
+ \,m_1\, m_3\, (\, \left(\rho _{13}+\rho _{23}-\rho _{12}\right) \partial_{\rho_{23}} + \left(\rho _{12}-\rho _{23}+\rho _{13}\right) \partial_{\rho_{12}} \,)  \bigg]
 \ , \nonumber
\end{aligned}
\end{equation}
\begin{equation}\label{}
\begin{aligned}
F_3^{(q)} \ = & \  (\rho_{12}^2+\rho_{13}^2+\rho_{23}^2-2 \left(\rho _{12} \rho_{13}+\rho _{23} \rho_{13}+\rho_{12} \rho _{23}\right))(\,m_1^2\partial^2_{\rho_{23}}-2m_1m_2\partial^2_{\rho_{23},\rho_{13}}+m_2^2\partial^2_{\rho_{13}} \,)
\\ &
\ + \ (d-1)\bigg[   m_2^2  \left(\rho_{13}-\rho_{12}-\rho _{23} \right)\partial_{\rho_{13}} \ + \  m_1^2 \left(\rho _{23}-\rho _{12}-\rho _{13}\right) \partial_{\rho_{23}}
\\ &
+ \,m_1\, m_2\, (\, (\rho _{23}+\rho_{12}-\rho_{13}) \partial_{\rho_{23}} + \left(\rho_{12}-\rho_{23}+\rho_{13}\right) \partial_{\rho_{13}} \,)  \bigg]
 \ , \nonumber
\end{aligned}
\end{equation}
commute with the free Hamiltonian $S_1^{(q)}$ as well as
\[
  L^{(q)}_0 \ = \  m_3\,[\,(m_1+m_2)\rho_{13}\ -\ (m_1+m_2)\rho_{23}\ +\ (m_1-m_2)\rho_{12}\, ]\,\partial_{\rho_{12}}\ +
\]
\[
  m_2\,[\,(m_1+m_3)\rho_{23}\ -\ (m_1+m_3)\rho_{12}\ +\ (m_3-m_1)\rho_{13}\,]\,\partial_{\rho_{13}}\ +
\]
\[
   m_1\,[\,(m_2+m_3)\rho_{12}\ -\ (m_2+m_3)\rho_{13}\ +\ (m_2-m_3)\rho_{23}\,]\,\partial_{\rho_{23}}\ .
\]
Furthermore, similar to the classical case, if the conditions (\ref{super-int}) are imposed,
the original 3-body quantum harmonic oscillator for $S-$states, described by the Hamiltonian
\begin{equation}
\label{HquS}
 {\cal H}^{(q)}\ = \  2\,S_1^{(q)} \ +\ V^{(ex)} \ \equiv
 \ -\De_{\rm rad}\ +\ \ 2\,\om^2\,\left(\nu_{12}\,\rho_{12}\ + \ \nu_{13}\,\rho_{13}\ + \ \nu_{23}\,\rho_{23}\right) \ ,
\end{equation}
(c.f. (\ref{HclaS})), is \emph{maximally} superintegrable. The triplet
$\{{\cal H}^{(q)},\,F^{(q)}_1,\,\tilde S^{(q)}_3\}$, where
\[
  \tilde S_3^{(q)} \ = \  S_3^{(q)} \ + \ \om^2\,\frac{m_1\nu_{13}}{m_3(m_1+m_2+m_3)} \bigg(m_2\,m_3\,\rho_{23} - m_2(m_2+m_3)\,\rho_{12} -m_3(m_2+m_3)\,\rho_{13} \bigg)\ ,
\]
spans a commutative Lie algebra. Also $F^{(q)}_2$ and $L^{(q)}_0$ commute with the Hamiltonian ${\cal H}^{(q)}$.
Note that if only the single condition (\ref{super-int-min})
\begin{equation*}
%\label{super-int-min}
      m_2\,\nu_{13}=m_3\,\nu_{12}\ ,
\end{equation*}
holds, then it can be shown that $F^{(q)}_2, F^{(q)}_3$ are not conserved and the quantum system ${\cal H}^{(q)}$ is {\it minimally} superintegrable: the triplet $(\tilde S^{(q)}_3, F^{(q)}_1, L^{(q)}_0)$ commutes with the Hamiltonian.

Now we proceed to the  question of variable separation. Following the general theory \cite{KKM:2018}, we  can  show that  separation of variables in the eigenvalue equation for free 3-body Hamiltonian $\De_{\rm rad}=-2\,S_1^{(q)}$ (\ref{addition3-3r-M}) occurs in the coordinates $\{w_1,\,w_2,\,w_3\}$,
\[
 w_1\ = \ \rho_{23}\quad ,
 \quad w_2\ = \ (m_2+m_3)\,m_3\,\rho_{13}+(m_2+m_3)\,m_2\,\rho_{12}-m_2\,m_3\,\rho_{23} \ ,
\]
\[
    w_3\ = \ \frac{\rho_{23} \big(\rho_{12}\,m_2\,(m_2+m_3) \,+\, \rho_{13}\,m_3\,(m_2+m_3)\,-\rho_{23}\,m_2\,m_3\big)}
   {\left[ (\rho_{23}-\rho_{13}+\rho_{12})m_2-m_3(\rho_{23}+\rho_{13}-\rho_{12}) \right]^2\,(m_3+m_2)}\
\]
\begin{equation}
\label{w123}
  \ =\ \frac{w_1 w_2}
   {\left[ (\rho_{23}-\rho_{13}+\rho_{12})m_2-m_3(\rho_{23}+\rho_{13}-\rho_{12}) \right]^2\,(m_3+m_2)}\ .
\end{equation}
%with inverse relations
%\[
%     \rho_{23}\ = \ w_1\ ,
%\]
%\[
%    \rho_{12}\ = \ \frac{(m_2+m_3)(\,m_3^2\,w_1 \, +\, w_2\,)\,w_3 \, \pm \, m_3\,\sqrt{(m_2+m_3)\,w_1\,w_2\,w_3} }{(m_2+m_3)^3\,w_3}\ ,
%\]
%\[
%   \rho_{13}\ = \ \frac{(m_2+m_3)(\,m_2^2\,w_1 \, +\, w_2\,)\,w_3 \, \mp \, m_2\,\sqrt{(m_2+m_3)\,w_1\,w_2\,w_3} }{(m_2+m_3)^3\,w_3} \  .
%\]
In these coordinates the (quantum) radial operator takes the form
\[
 \De_{\rm rad}\ = \ \frac{m_2+m_3}{m_2\,m_3}\left(2\,w_1\,\partial_{w_1}^2+d\,\partial_{w_1}\right) +\frac{(m_2+m_3)(m_1+m_2+m_3)}{m_1}\left(2\,w_2\partial_{w_2}^2\ +\ d\pa_{w_2}\right)
\]
\[
   +\ (m_2+m_3) \left(\frac{1}{m_2\,m_3\,w_1}+\frac{m_1+m_2+m_3}{m_1\,w_2}\right) \times
\]
\begin{equation}
\label{opham}
   \big[2\,w_3^2\,(\,4\,w_3(m_2+m_3)-1\,)\pa_{w_3}^2\,+\,w_3\big(12\,w_3(m_2+m_3)+d-4\big)
   \pa_{w_3}\big]\ .
\end{equation}
It is not algebraic anymore.

In order to demonstrate explicitly the separation of variables we consider the spectral problem for the third term in (\ref{opham}), involving the variable $w_3$ only,
\[
 \big[\,2\,w_3^2\,(\,4\,w_3(m_2+m_3)-1\,)\pa_{w_3}^2\,+\,w_3\big(12\,w_3(m_2+m_3)+d-4\big)
   \pa_{w_3}\,\big]\,\Theta \ = \ \la\,\Theta \ ,
\]
where $\la$ is a spectral parameter. Making now a gauge rotation of (\ref{opham}) with gauge factor $\Theta$,
\[
 \Theta^{-1}\,\De_{\rm rad}\,\Theta \ = \ \frac{m_2+m_3}{m_2\,m_3}\bigg[\ 2\,w_1\,\pa_{w_1}^2\ +\ d\,\pa_{w_1}  \ + \ \frac{\la}{w_1}\bigg] \ +
\]
\[
 \frac{(m_2+m_3)(m_1+m_2+m_3)}{m_1}\bigg[\ 2\,w_2\pa_{w_2}^2\ +\ d\,\pa_{w_2} \ + \  \frac{\la}{w_2} \bigg] \ ,
\]
we obtain an operator which depends on $w_{1,2}$ only in additive form and contains a type of effective potential. It admits separation of variables $w_1$ and $w_2$ in the standard way:
$a H(w_1;\la) + b H(w_2;\la)$ with eigenfunction in the form of the product $W(w_1) W(w_2)$, where the spectral problem
\begin{equation}
\label{Hw}
  H(w;\la) W(w) \equiv \bigg[\,2\,w\,\pa_{w}^2  \ + \ d\,\pa_{w}  \ + \ \frac{\la}{w}\bigg] W(w)\ =\ \vep\, W(w)\ ,
\end{equation}
defines the function $W$ and $\vep$ plays a role of the spectral parameter, $a,b$ are mass-dependent parameters. Thus, any eigenfunction of the free 3-body Hamiltonian $2 S_1^{(q)}$ has the form of the product $W(w_1) W(w_2) \Theta(w_3)$.

In $w$'s variables the harmonic oscillator potential (\ref{V3-es}) takes the form
\[
  V^{(ex)} \ = \ 2\,\om^2\,\left(\nu_{12}\,\rho_{12}\ + \ \nu_{13}\,\rho_{13}\ + \ \nu_{23}\,\rho_{23}\right)
\]
\[
   \ =  2\,\om^2\,\bigg[\,\frac{m_3^2\,\nu_{12}+m_2^2\,\nu_{13} +
   {(m_2+m_3)}^2\,\nu_{23}}{{(m_2+m_3)}^2}\,w_1 \ + \  \frac{\nu_{12}+\nu_{13}}{{(m_2+m_3)}^2}\,w_2 \,\pm\, \frac{m_3\,\nu_{12}-m_2\,\nu_{13}}{{(m_2+m_3)}^{\frac{5}{2}}}\,
   \sqrt{\frac{w_1\,w_2}{w_3}}  \, \bigg] \ .
\]
It is clear that if the condition $$m_2\,\nu_{13}\,=\,m_3\,\nu_{12}\ ,$$
see (\ref{super-int-min}), is imposed the potential becomes defined unambiguously and also becomes $w_3$-independent. Therefore, the 3-body quantum harmonic oscillator ${\cal H}^{(q)}$ (\ref{HquS}) written in $w$'s coordinates admits complete separation of variables. It is worth emphasizing that in this case the problem is {\it minimally} superintegrable.

\bigskip

{\bf Conclusions}

\bigskip

We defined a 3-body harmonic oscillator with pairwise interaction and showed that for $S$-states - the states with zero total angular momentum - in the  3-dimensional space of relative motion parametrized by squared relative distances, the problem has a hidden algebra $sl(4, {\bf R})$ and is exactly-solvable. The eigenvalues are linear in quantum numbers and the eigenfunctions are polynomials in three variables multiplied by a Gaussian function in relative distances. For $d=1$ a certain degeneracy occurs: the problem is reduced to a 2-dimensional one and the hidden algebra becomes $sl(3, {\bf R})$ acting in the space of relative distances $x_{ij}$.
We have exhibited a new $3d$ non-conformally flat oscillator system that is separable and maximally superintegrable. Almost all of the structure and classification theory for superintegrable systems applies only to conformally flat spaces, e.g. \cite{KKM:2018}. Examples on non-conformally flat spaces are relatively rare and thus valuable. The integrability results presented here were derived for arbitrary masses that obey no algebraic relations in general.  It is possible that for some special values of the masses and spring constants additional integrals appear.

A generalization to the general $n$-body system of interactive (an)harmonic oscillators in a $d$-dimensional space with $d>n-2$ is straightforward, while for smaller $d \leq n-2$ a certain complications occur: in general, the form of $\De_{\rm rad}$ is unknown. It will be considered elsewhere.

\section*{Acknowledgments}

A.V.T. is thankful to University of Minnesota, USA for kind hospitality extended to him where this work was initiated and IHES, France, where it was mostly completed.
W.M. was partially supported by a grant from the Simons Foundation (\# 412351 to Willard Miller, Jr.). M.A.E.R. is grateful to ICN-UNAM (Mexico) for kind hospitality extended to him where a part of this work was done during his numerous visits.
This research is partially supported by DGAPA IN113819 and CONACyT A1-S-17364 grants (Mexico).


\newpage

\begin{thebibliography}{99}

\bibitem{Delves:1960}
        L.M.~Delves,\\
        \textit{Tertiary and general-order collisions (II)},\\
        {\it Nucl.Phys. \bf 20}, 275-308 (1960)

\bibitem{Willard:2018}
        W.~Miller,~Jr., A.V.~Turbiner and M.A.~Escobar-Ruiz,  \\
        \textit{The quantum $n$-body problem in dimension $d\geq n-1$: ground state },\\
%        arXiv: 1709.01108v3, pp.32 (September 2017, updated: December 2017)
        {\it J. Phys. \bf A51} (2018) 205201 (25pp)
%              doi.org/10.1088/1751-8121/aabb10

\bibitem{Fortunato:2017}
        L.~Fortunato and T.~Oishi,\\
        \textit{Diagonalization scheme for the many-body Schr\"odinger equation},\\
        ArXiv:1701.04684v1, 9pp (January 2017)

\bibitem{Turbiner-Ush:1987}
        A.V.~Turbiner and A.G.~Ushveridze,\\
        \textit{Spectral Singularities and the Quasi-Exactly-Solvable Problem},\\
        {\it Phys.Lett. \bf 126A}, 181-183 (1987)

\bibitem{Turbiner:1988}
        A.V.~Turbiner, \\
        \textit{Quasi-Exactly-Solvable Problems and the $sl(2,R)$ algebra},\\
        \textit{Comm.Math.Phys. \bf 118} (1988) 467-474

\bibitem{Turbiner:2016}
        A.V.~Turbiner, \\
        \textit{One-dimensional Quasi-Exactly-Solvable Schr\"odinger equations},\\
        \textit{Phys. Repts. \bf 642} (2016) 1-71

\bibitem{TMA:2016}
        A.V.~Turbiner, W.~Miller,~Jr. and M.A.~Escobar-Ruiz,  \\
        \textit{Three-body problem in 3D space: ground state, (quasi)-exact-solvability},\\
%        IHES preprint P/16/29 (2016)\\
%        arXiv: 1611.08157v3, pp.24 (November 2016, updated: February 2017)\\
        {\it Journal of Phys. \bf A50} (2017) 215201 (19pp)

\bibitem{TME3-d}
         A.~Turbiner, W.~Miller,~Jr. and M.~A.~Escobar-Ruiz,\\
         \textit{Three-body problem in $d$-dimensional space: ground state, (quasi)-exact-solvability},\\
%         ArXiv: 1707.01324, pp.39 (July 2017)\\
         {\it Journ of Math Physics \bf 59} (2018) 022108 (29pp)

\bibitem{Fernandez}
         F.M.~Fern\'andez,\\
         \textit{Born-Oppenheimer approximation for a harmonic molecule},\\
          ArXiv: 0810.2210v2 , pp.14 (July 2017)\\


%\bibitem{Calogero:1969}
%         F.~Calogero,\\
%         \textit{Solution of a three-body problem in one dimension},\\
%         {\it Journ Math Physics \bf 10} (1969) 2191-2196

\bibitem{Ruhl:1995}
         W.~R\"uhl and A.V.~Turbiner,\\
         \textit{Exact solvability of the Calogero and Sutherland models},
         {\it Mod. Phys. Lett. \bf A10} (1995) 2213-2222, hep-th/9506105

\bibitem{KKM:2018}
        E.~G.~Kalnins, J.~M.~Kress and W.~Miller,~Jr.,\\ Separation of variables and Superintegrability: The symmetry of solvable systems,\\
        published by {\it Institute of Physics, UK}, 2018, ISBN: 978-0-7503-1314-8,
        309 pp.




\end{thebibliography}
\end{document}